\documentclass[manuscript, screen]{acmart}
\usepackage{booktabs} % For formal tables

\acmJournal{CSUR}
\acmYear{2020}
\setcopyright{acmcopyright}

\usepackage[english]{babel}
\usepackage[utf8x]{inputenc}
\usepackage{amsmath}
\usepackage{graphicx}
\usepackage{subcaption}
\usepackage{physics}
\usepackage{algorithm}
\usepackage[noend]{algpseudocode}
\usepackage[colorinlistoftodos]{todonotes}
\usepackage{xcolor}
\usepackage{multirow}

\begin{document}

\title{A Survey of Quantum Theory Inspired Approaches to Information Retrieval}
\titlenote{This work is funded by the European Union's Horizon 2020 research and innovation programme under the Marie Sklodowska-Curie grant agreement No 721321, and Natural Science Foundation of China (Grant No U1636203). Corresponding author: Dawei Song.}
\titlenote{This manuscript has been accepted for publication at ACM Computing Surveys on May 20, 2020}
\author{Sagar Uprety}
\affiliation{%
  \institution{The Open University}
  \country{UK}}
\email{sagar.uprety@open.ac.uk}
\author{Dimitris Gkoumas}
\affiliation{%
  \institution{The Open University}
  \city{Milton Keynes}
  \country{UK}
}
\email{dimitris.gkoumas@open.ac.uk}

\author{Dawei Song}
\affiliation{%
  \institution{Beijing Institute of Technology, China \& The Open University, UK}
\city{Milton Keynes}
 \country{UK}
}
\email{dawei.song2010@gmail.com}

\begin{abstract}
Since 2004, researchers have been using the mathematical framework of Quantum Theory (QT) in Information Retrieval (IR). QT offers a generalized probability and logic framework. Such a framework has been shown capable of unifying the representation, ranking and user cognitive aspects of IR, and helpful in developing more dynamic, adaptive and context-aware IR systems. Although Quantum-inspired IR is still a growing area, a wide array of work in different aspects of IR has been done and produced promising results. This paper presents a survey of the research done in this area, aiming to show the landscape of the field and draw a road-map of future directions.
\end{abstract}




\maketitle

\vspace{-2mm}
\section{Introduction}
Information Retrieval (IR) is the process of finding information that is relevant to the need of a user. The last two decades have completely changed how humans consume and interact with information. This change has been driven by the advances in web search engines, the ease of access to the Internet and the explosion of information available online. Information pertaining to a variety of needs is available - from lecture slides to news articles to descriptions and reviews of items, and so on. It becomes imperative that the IR systems continually improve to accommodate such information needs, which have been growing both qualitatively (in terms of complexity) and quantitatively. Essentially, the task of IR systems can be reduced to two aspects. One is how to efficiently and effectively represent and rank the variety of unstructured information being created at each instant. This involves tasks like indexing and improved understanding of the content through advanced representation methods, as well as ranking of the information items based on the representation. For example, representation of textual information can be improved with better understanding of natural language. The second is how to make an IR system better understand user's complex information need and information seeking behaviour. This involves understanding user's search context, search task and intent, and ability to measure task completion and user satisfaction through user interactions.

IR researchers have been investigating different approaches to improve IR systems from both the system point of view (representation and ranking) and the user point of view. Various areas in IR, for example, Neural IR~\cite{neural-information-retrieval-mitra, neuralIR_Onal2018}, Interactive IR~\cite{interactiveIR_Ruthven2009, interactiveIR_Borlund}, Cognitive IR~\cite{cognitiveIR_ingwersen, cognitive_IR_Sutcliffe1998, user_cognitive_IR}, and Dynamic IR~\cite{dynamic_ir_Sloan2015, dynamic_ir_grace}, have been developed. Quantum-inspired IR (QIR) is one such area, where the mathematical framework of Quantum Theory (QT) is utilized to develop representation and user models in IR that are expected to better align with human cognitive information processing. It is different from the field of Quantum Computing in that it does not involve computations based on physical quantum states. 

The benefits of using QT in IR are many-folds. It offers a new way of representing events and computing probabilities of events. Instead of the set-theoretic method of representing events as subsets of a larger sample space, QT represents events as subspaces of an abstract, complex vector space (called Hilbert space)~\cite{Busemeyer:2012:QMC:2385442}. Moreover, the same event can have multiple representations in multiple basis of the Hilbert space. This method of representation can help in the abstraction and contextualization of information objects like documents and queries~\cite{Rijsbergen:2004:GIR:993731}. For example, if a set of basis vectors correspond to a set of documents, another set of basis vectors in the same Hilbert space can represent the same set of documents in a different context. Hence a query (as an event) will be represented by these different basis depending upon the context of retrieval. The Hilbert space representation of events also leads to a generalized method of calculating probabilities (Born rule)~\cite{Born1926}, by taking into account interference between events. This can model a user's decisions under ambiguity better than traditional probability models~\cite{Busemeyer:2012:QMC:2385442}. Such a representation method can inherently model incompatible variables - those where measurement on one variable affects the outcome of the other. For two such incompatible variables $A$ and $B$, measuring $A$ would alter the state of the system, so that the subsequent measurement of $B$ would be different than if it was measured alone or before $A$. Thus these two variables cannot be measured simultaneously or jointly, and different orders of measurement would lead to different outcomes. Traditional probability theory assumes that for any pair of events, $p(A,B) = p(B,A)$, which would be incorrect for incompatible variables. The cognitive phenomenon of order effect is generally considered to be a consequence of incompatibility in measuring human decisions~\cite{Busemeyer:2012:QMC:2385442}. There has been a lot of research in recent years, which shows the presence of Order Effects in relevance judgment of documents (detailed in Section 3).

\begin{figure}
    \centering
    \includegraphics[width=\textwidth]{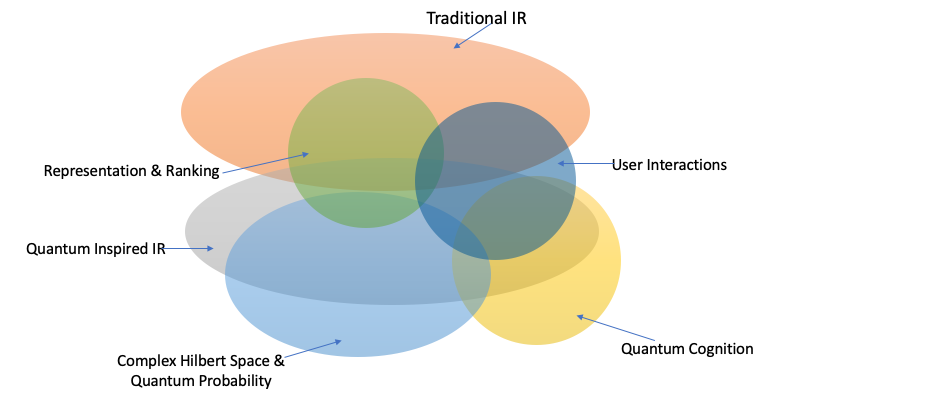}\vspace{-4mm}
    \caption{Brief overview of quantum-inspired IR}
    \label{quantum_ir_brief}
    \vspace{-4mm}
\end{figure}

Correspondingly, the application of QT to IR can be broadly divided into two subareas: (1) Representation and Ranking, and (2) User Interaction. Figure \ref{quantum_ir_brief} shows a sketch of the overlap between traditional IR and Quantum-inspired IR, and their underlying components. We show traditional IR in terms of the two sub-areas, which overlap because user interactions like relevance feedback are often used in re-ranking tasks. In this sense, QIR overlaps with traditional IR as it is also divided into these two sub-areas. The difference comes in the tools used by QIR. It utilises the mathematical framework of QT including complex Hilbert space models for representation learning and quantum probability rules to model cognitive interference in document ranking. 

QIR borrows heavily from  concepts, models and techniques developed in the field of Quantum Cognition, especially in the modelling and incorporating the user interactions in IR. QT has been successfully applied to model and predict irrational human decision making and explain cognitive biases in human judgments~\cite{Pothos2011AQP,Pothos2013,Busemeyer2011-conjunction,Wang2013,Trueblood2011,Pothos2009}. The emerging field of Quantum Cognition~\cite{Busemeyer:2012:QMC:2385442} studies such quantum-like phenomena in cognitive and decision sciences. There is already a growing community of researchers, under the umbrella of Quantum Interaction (see http://www.quantuminteraction.org/home), who are applying QT to various disciplines such as Biology, Cognition, Economics, Natural Language Processing and Information Retrieval. In 2017, a major project which seeks to investigate a Quantum Theoretical approach to IR (QUARTZ - see http://www.quartz-itn.eu/) has started, under the Marie Skłodowska Curie Actions scheme of the European Union's Horizon 2020 programme, with 7 participating universities all over Europe and several external partners around the world. 

QIR is a growing multi-disciplinary area and has been attracting an increasing attention of researchers in IR. Especially, the recent several years have witnessed a large number of models and applications of QIR which have shown good results and a great potential. However, the field lacks a comprehensive review of the literature, and the individual works are largely segmented. This is why a survey paper is urgently needed. It is important and timely to review the literature systematically to provide a clear picture of the landscape and a road-map for the future. Although this is not the first work to accumulate findings in QIR. Around ten years ago, a position paper~\cite{Song2010HowQT}, organised QIR into three themes: frameworks, spaces, and interference. However, it was more on a conceptual level and the field of QIR was in its infancy. After a decade of development since then, the landscape of QIR research has significantly changed. A large number of more comprehensive and larger scale QIR approaches have been developed covering different aspects of IR, and have achieved remarkable experimental results. 

The next section introduces the basic concepts and notations of QT, to enable readers to understand their usage in IR. Section 3 reviews the literature of Quantum-inspired IR, followed by a discussion on its shortcomings and benefits in Section 4. Section 5 concludes the paper by discussing future work directions in Quantum-inspired IR. 

\section{Quantum Theory Preliminaries}

Quantum Theory is also regarded as a theory for calculating probabilities~\cite{pitowsky2006quantum}, which was developed in the first half of the twentieth century to explain the counter-intuitive probabilistic outcomes of experiments on microscopic particles. These results could not be explained using standard probabilistic models. Quantum Mechanics was later axiomatically organized by John von Neumann~\cite{vonNeumann}, thus enabling it to be used as an abstract mathematical framework even outside of Physics. The fundamental difference between the classical and quantum probabilities lies in the representation of events. In the classical probability theory, events are represented as subsets of a sample space. In the quantum probability theory, events are represented as subspaces of an abstract vector space. As such, the quantum probability theory is a generalization of the classical probability theory, and can be useful in calculating the probabilities of events which cannot be represented in a set-theoretic formalism due to their inherent structure. The use of QT for applications beyond Physics was first suggested by Niels Bohr~\cite{bohr-complementarity} (pg. 294-295, 297), one of the founding fathers of Quantum Mechanics. He mentioned the existence of complementary variables in Psychology as similar to the incompatible properties of quantum systems. As we will discuss in this section, QT provides a method to model incompatible variables naturally. In the sub-sections to follow, a brief description of the need for the quantum probabilistic framework is provided and the formal concepts underlying QT are discussed.

\vspace{-2mm}
\subsection{The Double Slit Experiment}
\vspace{-1mm}
\begin{figure}[t!]
\begin{center}
\includegraphics[width=0.8\textwidth]{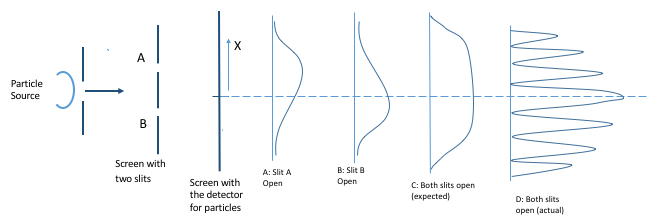}
\end{center}
\vspace{-2mm}
        \caption{Double slit experiment setup}
        \label{double-slit-1}
        \vspace{-4mm}
\end{figure}

The earliest experiment on microscopic particles which puzzled physicists was the Double Slit Experiment. Consider Figure \ref{double-slit-1}, in which microscopic particles, say electrons, are fired from a source to a screen consisting of two slits. On the right of this screen is another screen made up of detectors, which can detect the arrival of a particle as a function of its distance $x$ from the center of this screen. By measuring the mean number of pulses, one can measure the probability of the electron reaching the detector screen as a function of $x$. When only one slit is open, the probability distribution obtained looks like the one in Figure \ref{double-slit-1}(A) and \ref{double-slit-1}(B) for slits A and B respectively. Now, according to the classical probability theory, opening both slits at the same time should lead to the sum of the probability distributions as shown in Figure \ref{double-slit-1}(C). However, it was actually seen that on opening both slits, one obtains a distribution of electrons as \ref{double-slit-1}(D). It is a complicated curve having several maxima and minima, indicating that there are some locations on the detector screen that electrons never register. This distribution is the same as obtained in the case of interference of waves. The interference pattern is produced by adding the amplitudes of two waves and the squaring the sum to get the intensity. Hence the data of curve \ref{double-slit-1}(D) can be explained by assuming that the electrons behave like waves when traveling from the source through the slits to the detector screen. In doing so, it is as if a single electron goes through both the slits at the same time - a fundamental quantum property called \emph{superposition}. A complex number called the probability 
amplitude is ascribed to the electron corresponding to the two possible paths. Let $\phi_a$ be the probability amplitude for the path $S \rightarrow A \rightarrow X$ and $\phi_b$ be the probability amplitude of the electron for the path $S \rightarrow B \rightarrow X$, when the slits $A$ and $B$ are opened respectively, $S$ being the source.

 The amplitudes differ because of the difference in the complex phase for the two paths taken. The probabilities are calculated, according to the Born rule~\cite{Born1926}, as the square of the amplitudes. Thus the probability of detecting an electron at a position $X$ from the center of the detector screen, when only slit $A$ is open, is $p(A) = |\phi_a|*|\phi_a^\dagger| = |\phi_a|^2$ (where $\phi_a^\dagger$ is the complex conjugate of $\phi_a$). When both slits are open, the probabilities are calculated by following the Law of Total Amplitude~\cite{feynman-3}. The probability amplitudes for the two paths are added up and then the probability is calculated by taking the square of the sum: 
 \vspace{-2mm}
\begin{align} \label{interference-term}
p(X) &= |\phi_a + \phi_b|^2 \\ \nonumber
	 &= |\phi_a|^2 + |\phi_b|^2 + 2|\phi_a|*|\phi_b| \\ \nonumber
     &= p(A) + p(B)+ 2\sqrt{p(A)}\sqrt{p(B)}\cos(\theta)
\end{align}
where $\theta$ is the phase difference between the two paths. The negative values of $\cos{\theta}$ are responsible for the minima obtained in the curve \ref{double-slit-1}(D). For $\theta = \frac{\pi}{2}$, we get the classical probabilities as a special case.

Thus we see that the origin of quantum probabilities lies in the Law of Total Amplitude and the Born rule. When a quantum entity can take one or more paths, it takes all of them at the same time, and the quantum entity is said to be in a \emph{superposition state} of all possible paths. These paths influence each other, in a phenomenon called \emph{quantum interference}, which gives rise to the extra terms in the calculation of probabilities.

It should be noted that when we say quantum probabilities, the concept of probability remains the same as classical probabilities. To paraphrase Richard Feynman, ``If the probability of a certain outcome of an experiment is $p$, then if the experiment is repeated many times one expects that the
fraction of those which give the desired outcome is $p$. What changes in QT is only the method of calculating probabilities''~\cite{feynman-3}.

\subsection{The Axioms of Quantum Theory}

\subsubsection{Representation of Events} \label{rep_of_events}
QT provides a new method of assigning probabilities to events. In the classical method of calculating probabilities, we assume a finite sample space consisting of $N$ points. The collections of all the points in the space is described as a set $X = \{x_1, x_2, ..., x_N\}$. An event is any subset of $X$, say $A\subseteq X$. For two such events $A\subseteq X$ and $B\subseteq X$, $A\cup B$ and $A\cap B$ are also events. Atomic events are given by singletons. 

Instead of the sample space of events, a complex Hilbert space of infinite dimensions is used in QT. A Hilbert space is an abstract vector space, which includes a complex inner product between any two vectors in the space. For simplicity, we deal with a finite dimensional Hilbert space here. A $N$-dimensional Hilbert space comprises $N$ orthonormal basis vectors $X = \bigl\{\ket{X_i}, i = 1, ..., N\bigr\}$. The choice of basis is arbitrary and there can be any number of basis for a Hilbert space. Here $\ket{X}$ is the way to denote a vector in the Dirac notation~\cite{dirac}. An event $A$ is defined not by the subset of vectors $X_A \subseteq X$, but rather by a subspace spanned by this subset. If $A$ is an event spanned by $X_A \subseteq X$ and $B$ is an event spanned by $X_B \subseteq X$, the intersection of the two events, also called the "meet" and denoted as $A \wedge B$, is given by the span of vectors in the subset $X_A\cap X_B$. Similarly, the union of the events, called the "join" and denoted as $X_A \vee X_B$, is given by the span of vectors in $X_A\cup X_B$. Note how the set theoretical intersection and union of points are replaced by the span of the intersection and union of vectors. This structural property leads to the violation of the distributive axiom~\cite{Busemeyer:2012:QMC:2385442}. Before talking about that further, we first discuss the concept of states and projectors in the quantum framework.

\subsubsection{States of a Quantum System}
In the classical framework, we have the concept of a probability distribution function $p(X_i)$, which assigns real numbers to each point $X_i$ of a sample space. In the quantum framework, we define a state vector $\ket{S}$ of unit length in a Hilbert space $X$, which induces a probability distribution over the subspaces of the Hilbert space (Figure \ref{sample-hilbert-space-a}). A subspace is represented in term of a projection operator $P$, which is Hermitian ($P\dagger = P$, where $P\dagger$ denotes the complex conjugate of the transpose of $P$) and Idempotent ($PP = P$). The probability induced by a state vector $\ket{S}$ onto a subspace is given by the square of the projection of the vector onto the subspace. It is calculated as:
\vspace{-2mm}
\begin{align}
|P\ket{S}|^2 &= \bra{S}P^\dagger P\ket{S} \\ \nonumber
			 &= \bra{S}P\ket{S}
\end{align}

Figure \ref{sample-hilbert-space}(b) shows a two-dimensional Hilbert space with state vector $\ket{S}$ projected onto a one-dimensional subspace, $A_2$. In this case the projector is given by $P_{A_2} = \ket{A_2}\bra{A_2}$ and the probability distribution of the state given by the vector $\ket{S}$ is:
\vspace{-4mm}
\begin{align}\label{projection-prob}
|P_{A_2}\ket{S}|^2  &= \bra{S}P_{A_2}\ket{S} \\ \nonumber
 &= \bra{S}\ket{A_2}\bra{A_2}\ket{S} \\ \nonumber
 &= |\bra{A_2}\ket{S}|^2
\end{align}

Here the quantity $\bra{A_2}\ket{S}$ is the probability amplitude of the state $\ket{S}$ for the event $A_2$. The state of a quantum system $\ket{S}$ is in general a superposition of all possible events (see the next subsection for a discussion on superposition). As discussed before, the events are given by all the vectors of an orthonormal basis. In the basis $\bigl\{\ket{A_1}, \ket{A_2} \bigr\}$, the state of the system is represented as:
\vspace{-4mm}
\begin{align}
\ket{S} = a_1\ket{A_1} + a_2\ket{A_2}
\end{align}
where $a_1 = \bra{A_1}\ket{S}$ and $a_2 = \bra{A_2}\ket{S}$ are the probability amplitudes and $|a_1|^2$, $|a_2|^2$ represent the probabilities for events $A_1$ and $A_2$ to occur for the state $\ket{S}$. Hence $|a_1|^2 + |a_2|^2 = 1$.

\subsubsection{Superposition and Collapse of a Quantum State}
In models based on the classical probability theory, like Bayesian networks, a state of a system evolves from ``moment to moment, but any given point of time the system is in a definite state''
~\cite{Busemeyer:2012:QMC:2385442}. To deal with uncertainty about which state the system is in, probabilities are assigned to each state. Thus, a dynamic system is in a definite state at each point of its evolution and is governed by a probability distribution over the states. 

A quantum system is different from classical systems because of its ability to be in a superposition of all the possible states at the same time. This superposed state is a new state, which is not the same as any of the possible states of a classical system. Rather, it encapsulates the possibilities of being in all possible states. When a measurement is performed on a quantum system to learn its state, the superposed state collapses into one of the possible states with a certain probability. As an example, the system with state $\ket{S}$ in Figure \ref{sample-hilbert-space}(a) is a superposition of the basis vectors $\ket{A_1}$ and $\ket{A_2}$. It is neither in state $\ket{A_1}$ nor in $\ket{A_2}$. It is a new state with probabilities $|a_1|^2$ and $|a_2|^2$ for the system to be in state $\ket{A_1}$ or $\ket{A_2}$ upon measurement. Changing the probability amplitudes $a_1$ and $a_2$ leads to a new state, different from $\ket{S}$. This concept of collapse of a superposed state into one of the constituent states upon measurement is referred to as the Copenhagen Interpretation of Quantum Theory.

\subsubsection{Violation of Distributive Axiom}
\begin{figure}
\centering
    \begin{subfigure}[t]{0.25\textwidth}
        \includegraphics[width=\textwidth]{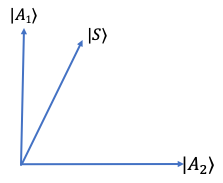}
        \caption{}
        \label{sample-hilbert-space-a}
    \end{subfigure}
    \begin{subfigure}[t]{0.35\textwidth}
        \includegraphics[width=\textwidth]{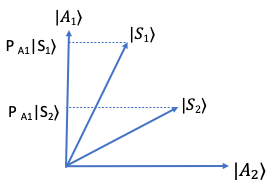}
         \caption{}
         \label{sample-hilbert-space-b}
  \end{subfigure}
        \caption{A two dimensional Hilbert Space with initial state vector and its projection}
        \label{sample-hilbert-space}
\end{figure}  
In the classical probability theory, for a sample space $X = \bigl\{A, B \bigr\}$, the distributive axiom states that ``$A = A \cap (B \cup \widetilde{B}) =  (A \cap B) \cup (A \cap \widetilde{B})$, where $\widetilde{B}$ is the complement of event $B$. This axiom leads to the law of total probability''~\cite{Busemeyer:2012:QMC:2385442}:
\vspace{2mm}
\begin{align}
p(A) &= p(A \cap X) = p(A \cap(B \cup \widetilde{B})) \\ \nonumber
&= p((A \cap B) \cup (A \cap \widetilde{B})) \\ \nonumber
&= p(A \cap B) + p(A \cap \widetilde{B}) \\ \nonumber
&= p(B)p(A|B) + p(\widetilde{B})p(A|\widetilde{B})
\end{align}

In simple terms, this law states that if an event $A$ occurs, it can occur along with $B$ or without $B$. In the quantum framework, consider a two dimensional Hilbert space with two basis vectors $\ket{A_1}$ and $\ket{A_2}$, as in Figure \ref{sample-hilbert-space-a}. The intersection and union of two subspaces (called as $meet$ and $join$, respectively) is defined as the intersection and union of the set of vectors spanning the subspaces, respectively (Section \ref{rep_of_events}). Denoting the one-dimensional subspaces of this Hilbert space by their projectors $P_S, P_{A_1}, P_{A_2}$, the meet of the subspaces $P_S \wedge P_{A_1} = 0$ and $P_S \wedge P_{A_2} = 0$. The meet of two subspaces thus works the same way as intersection in set theory. The difference comes from the definitions of union and complement. The union or join of the two subspaces $P_{A_1} \vee P_{A_2}$ is the whole two-dimensional Hilbert Space, not just the set of two vectors $\ket{A_1}$ and $\ket{A_2}$ as in the set theory. Thus we get
\vspace{-2mm}
\begin{equation}
P_S \wedge (P_{A_1} \vee P_{A_2}) = P_S
\end{equation}
which violates the distributive axiom, as $(P_S \wedge P_{A_1}) \vee  (P_S \wedge P_{A_2}) = 0$.

\subsubsection{Compatible and Incompatible Events}

\begin{figure}
\centering
    \begin{subfigure}[t]{0.35\textwidth}
        \includegraphics[width=\textwidth]{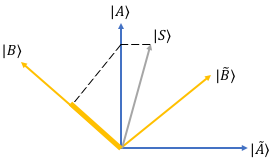}
        \caption{}
        \label{incompatible-a}
        \end{subfigure}
    \begin{subfigure}[t]{0.35\textwidth}
        \includegraphics[width=\textwidth]{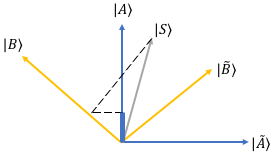}
        \caption{}
         \label{incompatible-b}
  \end{subfigure}
          \caption{Two basis representing incompatible events showing order effects}
         \label{incompatible}
         \vspace{-4mm}
\end{figure}

Classical systems follow the principle of unicity~\cite{Griffiths2001}, which states that there is always a single sample space ``which provides an exhaustive description of all the events that can happen in an experiment''~\cite{Busemeyer:2012:QMC:2385442}. Therefore a single probability distribution function is sufficient to calculate the probabilities for all the events. 

In the quantum framework, a state vector is represented as a superposition of all the basis vectors. One can choose to represent this state vector in any arbitrary basis. Thus the same state vector is expressed in different basis and each basis represents a particular property of the quantum system. The state vector induces different probabilities onto different basis of the Hilbert space. The state vector is thus an abstract entity. It does not have any fixed representation. A particular representation conceptualizes when we talk of a particular basis. 

In Figure ~\ref{incompatible}, we show a Hilbert space with two basis. One with orthonormal vectors $\ket{A}$ and $\ket{\widetilde{A}}$ and another basis with orthonormal vectors $\ket{B}$ and $\ket{\widetilde{B}}$. Consider the following events, in a particular order - $A$ and $B$. To calculate the probability that these two events occur, the state vector $\ket{S}$ is projected onto the vector $\ket{A}$ and the new collapsed state is projected onto the vector $\ket{B}$. Hence we get the probability for $A$ and $B$ to occur as $p(A,B) = |P_BP_A\ket{S}|^2$, and using Equation \ref{projection-prob}, we get 
\vspace{-2mm}
\begin{equation} \label{a,b}
p(A,B) = |\bra{B}\ket{A}|^2.|\bra{A}\ket{S}|^2
\end{equation}

Now, if the same two events occur in the reverse order, i.e., $B$ and then $A$, then the probability of them occurring is given by $p(B,A) = |P_AP_B\ket{S}|^2$, which, using Equation \ref{projection-prob}, is
\vspace{-2mm}
\begin{equation}\label{b,a}
p(B,A) = |\bra{A}\ket{B}|^2.|\bra{B}\ket{S}|^2
\end{equation}

Now, Equations \ref{a,b} and \ref{b,a} will assign different values to the left-hand side when the value of the terms $\bra{B}\ket{S}$ and $\bra{A}\ket{S}$ are different. Which is the case if $A$ and $B$ are vectors in different basis. In the classical theory, we can assign joint probability distributions to two events occurring together regardless of their order, i.e., $p(A,B) = p(B,A)$, but such a joint probability distribution does not exist for events belonging to two different basis in a Hilbert space. We call these events as \emph{incompatible events}. In the language of linear algebra, the projectors corresponding to these events do no commute, i.e., $P_AP_B \neq P_BP_A$. A geometrical explanation can be obtained from Figure~\ref{incompatible}. In Figure~\ref{incompatible-a}, the order of projections is $S \rightarrow A \rightarrow B$ and in Figure ~\ref{incompatible-b}, it is $S\rightarrow B \rightarrow A$. We can see that the final projections (indicated by thick blue lines) in the two cases are different (and hence different probabilities) and depend upon the geometry of the Hilbert space (specifically, the angle between the vectors).

\subsubsection{Density Matrices and Trace Rule}
Another way of representing a quantum state, apart from the vector representation, is the density matrix or the density operator $\rho$. For a state $\ket{\psi}$, the density matrix is given by
\vspace{-2mm}
\begin{equation}
\rho = \ket{\psi}\bra{\psi}
\end{equation}
which is a square matrix. The probability induced by the state represented by $\rho$ onto a subspace represented by the projector $P$ follows from the Gleason's Theorem~\cite{Gleason1957}, and is given by
\vspace{-2mm}
\begin{equation}
Pr = tr(\rho P)
\end{equation}
where $tr(x)$ is the trace of a matrix $x$, i.e. the sum of its principle diagonal elements. If we denote $P= \ket{\phi}\bra{\phi}$, then the trace can be written as 
$tr(\rho P) = tr(\rho \ket{\phi}\bra{\phi}) = \bra{\phi}\rho \ket{\phi}$ which, for $\rho = \ket{\psi}\bra{\psi}$ is
\vspace{-2mm}
\begin{equation}
\tr(\rho P) = |\bra{\psi}\ket{\phi}|^2
\end{equation}
This is the same probability as calculated using the Born rule described earlier.

Density Matrix gives us the advantage of representing a mixture of classical and quantum systems. For example, if there is a mixture of $n$ quantum systems, with the probability of each system being present in the mixture denoted by $p_i$, then this mixed system can be represented by a density matrix:
\vspace{-2mm}
\begin{equation}
\rho = p_1\rho_1 + p_2\rho_2 + ... + p_n\rho_n
\end{equation}
where $\rho_i$ is the density matrix of the $i$-th quantum system, which has a classical probability $p_i$ to belong to the mixture.

\subsubsection{Composite Quantum Systems}
Multiple quantum systems can be considered as a single system by combining their Hilbert spaces using a tensor product of the individual Hilbert spaces. If $\ket{S}_1, \ket{S}_2, ..., \ket{S}_n$ represent the states of $n$ distinct quantum systems, the state of the composite quantum system of all these individual systems is given by $\ket{S}_1 \otimes \ket{S}_2 \otimes ... \otimes \ket{S}_n$, also denoted as $\ket{S}_1\ket{S}_2..\ket{S}_n$. For example, consider two quantum systems represented by two dimensional Hilbert spaces with $\ket{A}$ and $\ket{\Tilde{A}}$ as the basis vectors. Note that this is different from the system represented in Figure~\ref{sample-hilbert-space}(b). Instead of multiple states in one Hilbert space, here we have multiple Hilbert spaces each with a state $\ket{S}_i$. The state of each of the systems (identical Hilbert spaces in this case) are given by:
$
\ket{S}_1 = \frac{1}{\sqrt{2}}\ket{A} + \frac{1}{\sqrt{2}}\Tilde{\ket{A}}
$ and $
\ket{S}_2 =  \frac{1}{\sqrt{2}}\ket{A} + \frac{1}{\sqrt{2}}\Tilde{\ket{A}}
$.
Then the composite system is given by: 
\begin{equation}
\vspace{-2mm}
\ket{S}_1 \otimes \ket{S}_2 = \frac{1}{2}(\ket{A}\ket{A} + \ket{A}\Tilde{\ket{A}} + \Tilde{\ket{A}}\ket{A} + \Tilde{\ket{A}}\Tilde{\ket{A}})
\end{equation}
The above composite state is a separable state. It can be factorized into two separable components as: \\ $(\frac{1}{\sqrt{2}}\ket{A} + \frac{1}{\sqrt{2}}\Tilde{\ket{A}}) \otimes 
(\frac{1}{\sqrt{2}}\ket{A} + \frac{1}{\sqrt{2}}\Tilde{\ket{A}})$. 
There exists some composite systems for which it is not possible to separate the composite state back into single systems. A famous example of such states are Bell states:
\vspace{-2mm}
\begin{align}
\ket{S^{\pm}} = \frac{1}{\sqrt{2}}(\ket{A}\Tilde{\ket{A}} \pm \Tilde{\ket{A}}\ket{A})
\end{align}
These states are called Entangled states and this property of Entanglement is a unique and a fundamental feature of Quantum Physics. When a measurement is performed on one part of the entangled system, the state of the other system can be known instantaneously, even if the two individual components are separated by a large distance. For example, consider two experimenters Alice and Bob who possess quantum states which are entangled with each other: $\ket{S} = \frac{1}{\sqrt{2}}(\ket{A}_1\Tilde{\ket{A}}_2 + \Tilde{\ket{A}}_1\ket{A}_2)$, where subscripts $1$ and $2$ denote that the states are possessed by Alice and Bob respectively. Now initially both the systems are in a superposition state. One cannot tell if it is in state $\ket{A}_i$ or state $\Tilde{\ket{A}}_i$ ($i \in {1,2})$. If Alice measures her system and it collapses to, say, state $\ket{A}_1$ (it can collapse to either $\ket{A}_1$ or $\Tilde{\ket{A}}_1$ with equal probability), then the state of the composite system collapses to state $\ket{A}_1\Tilde{\ket{A}}_2$. Alice can instantaneously know that Bob's state has collapsed to state $\Tilde{\ket{A}}_2$. 
\section{Quantum Theory inspired Information Retrieval}
\begin{figure}
\centering
\includegraphics[width=1.0\textwidth]{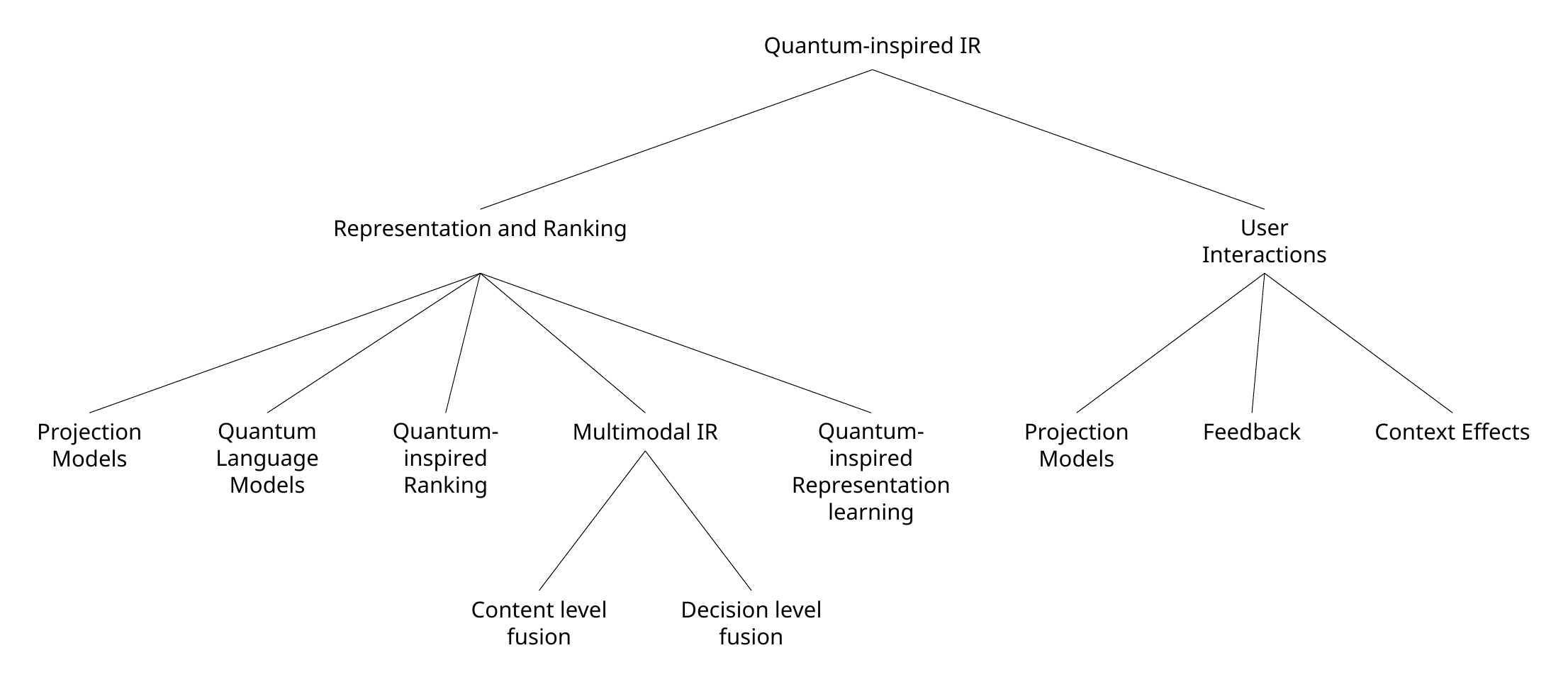}
        \caption{Structure of the Quantum inspired IR survey}
\label{QIR_overview}
\end{figure}

The application of QT to Information Retrieval (IR) can be broadly divided into two major aspects. The first is the Quantum-inspired representation of entities like documents, queries, etc. in IR. Related to the representational aspect is that of Ranking in IR. The way the documents and queries are represented often determines the method for ranking documents. The second aspect is User interactions, including relevance feedback, query expansion, and user cognitive modeling. Projection models using the Hilbert space and multiple basis to represent states are used in both aspects: in representation - for abstraction of documents and queries, and in modeling user's dynamic and contextual information needs. Representation and user interaction areas in Quantum-inspired IR can be further sub-divided based on the specific approaches and applications (Figure \ref{QIR_overview}).

\subsection{Representation and Ranking}

\subsubsection{Subspace Representation and Projection Models}

The first ideas regarding using the mathematical framework of QT in IR were described in the van Rijsbergen's pioneering book, \textit{The Geometry of Information Retrieval}~\cite{Rijsbergen:2004:GIR:993731}. It addresses the need to develop a formal theory unifying different IR models, namely logic, vector space, and probabilistic models. It also sought to explore a formal description of user interactions and the abstraction of the concept of a document in IR. Hence we find that the user is at center of most of the Quantum-inspired models and user interactions permeate all the representation and ranking methodologies, which we will discuss in the rest of this subsection.
\\
\\
\noindent\textit{Subspace Representation} 
\\
A representation of document is usually related to the text it contains, but a document is in general a more abstract entity. To quote~\cite{Rijsbergen:2004:GIR:993731}, ``it is a set of ideas, a set of concepts, a story, etc.'' A document is defined as an abstract object that encapsulates answers to all possible queries. This is similar to a state vector in QT, which encodes information about all possible outcomes of measurement. The user interaction with an IR system is considered akin to measurement in QT, and the abstract document materializes to the specific information need of the user upon interaction. The Hilbert space representation of the Quantum framework is utilized to represent documents and queries in IR. It might seem similar to the Vector Space Model (VSM). However, instead of modeling documents as vectors in a term space, they are represented as subspaces of a concept space, spanned by a set of basis vectors. 

Note that the documents and queries are themselves abstract and are defined in terms of the choice of basis. The same query or documents can be defined in different basis depending upon the user's point of view. The existence of multiple basis for the same state vector causes abstraction of objects in a Hilbert space. This, coupled with the fact that documents and queries are not merely vectors but subspaces in a complex, infinite dimensional vector space, gives us the leverage over the classical Vector Space Model. Besides providing a theoretical modification of the representational concepts of traditional IR, ~\cite{Rijsbergen:2004:GIR:993731} also shows how the existing IR tasks like co-ordination level matching, feedback, clustering, etc. can be performed using the Quantum-like formulation.

Modeling queries and documents as multiple basis in IR was also investigated in ~\cite{Melucci2005CanVS}. Documents and queries are modeled using certain semantic descriptors. However the semantic descriptors used for the same query or document may be different for different users, or different for same user in a different time, location or need. Therefore the use of descriptors depend upon the context. Since descriptors are modeled as basis vectors in a VSM, one can extend the VSM to include multiple basis, where each basis corresponds to a context. \cite{MelucciCMD} provides a method to discover different contexts from data to model them as different basis, using a matrix decomposition algorithm (i.e., Cholesky's decomposition). 
\\
\\
\noindent\textit{Information Need Spaces} 
\\
A further development of the Quantum-inspired IR paradigm~\cite{Piwowarski:SIR} advocates the use of an information need space to model user interaction and evolving information need (IN) as part of representation. An IN is represented as a state in the form of a density matrix. For an ambiguous needs, the state is a mixed state, and if the IN is completely specified, it is a pure state. Before any user interaction, the IR system starts as a mixed state of all possible IN states. Consider the example when a user wants to order a pizza. In the beginning, the IN is in a mixture of all possible states, but a query “pizza” restricts the information need space to a subspace. Further interactions like knowing the time of the day, location of the user, etc. leads to smaller subspaces. Hence the evolution of information need is captured in the geometry. The representation of documents is proposed as in Structured Information Retrieval (SIR), which breaks away from representing the whole document as a single retrieval unit and uses document fragments like sections or paragraphs in response to a user query. It has been shown in \cite{Piwowarski:2008:SCR:1416950.1416951} that answers to queries may correspond to document fragments and not the whole document.

The specific details of building the information need spaces are given in \cite{Piwowarski2010FilteringDW}. The documents are modeled as a set of INs, with each IN being a vector. Using the SIR approach, documents are divided into fragments - paragraphs, sentences, sections or the document itself. Each document is converted into a vector using traditional techniques like tf-idf. Each of these fragments can satisfy an information need. Further, spectral decomposition of this set of vectors is performed to construct the document subspace. If the set of vectors for a document is $U_d$, then a subspace $S_d$ comprises the span of the eigenvectors of the matrix $\sum_{u \in U_d}{uu^T}$. Only the eigenvectors corresponding to the top $k$ eigenvalues are considered, since the low eigenvalues can be associated with noise.

\begin{figure}
\centering
\includegraphics[width=0.58\textwidth]{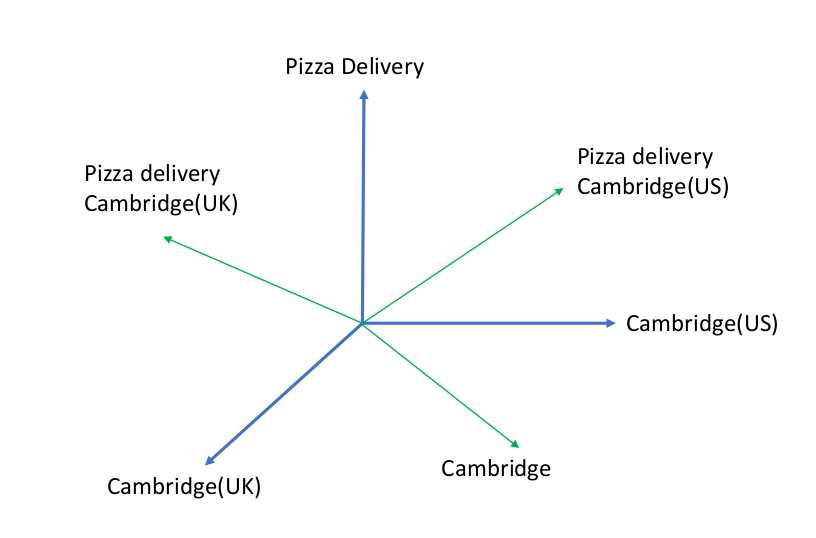}
        \caption{Three dimensional Information Need (IN) space}
\label{representation-query-doc}
\end{figure}

\cite{Piwowarski:2010:EMR:1937055.1937067} extend this work to include representation for queries. As a document is represented as a ``set of pure IN vectors corresponding to different fragments of the document, a query term $t$ is represented as a set $U_t$ of IN vectors that correspond to document fragments containing the term $t$''~\cite{Piwowarski:2010:EMR:1937055.1937067}. 
Consider two documents $D_1$ and $D_2$ consisting of three different paragraphs each. Let $U_1 = \{v_1, v_2, v_3\}$ and $U_2 = \{v_4, v_5, v_6\}$ be the IN vectors corresponding to the documents. Taking the simpler case of a single term query, suppose the term occurs in paragraphs corresponding to the vectors $v_2, v_5, v_6$. Assuming that each fragment is equally likely to be a pure IN and a part of the user's actual IN, the density matrix for the query is written as:
\vspace{-1mm}
\begin{equation} \label{query-density}
\rho_q = \frac{1}{N_t}\sum_{\varphi \in U_t}{\varphi\varphi^T}
\end{equation}
where $N_t = 3$ is the number of document fragments a term occurs in. Denoting the $S_d = \sum_{u \in U_d}{uu^T}$ as projector for a document as explained above, we can calculate the probability of relevance of the document for the query as: 
\vspace{-2mm}
\begin{equation}
p(Rel|q,d) = tr(\rho_qS_d)
\end{equation}
For queries with multiple terms, either a weighted mixture of density matrices for the terms, or in an interesting case, the density matrix for a superposition of pure IN vectors can be used. Consider the three dimensional subspace of an information need space as shown in Figure \ref{representation-query-doc}. Let the vectors $\varphi_p, \varphi_{uk}, \varphi_{us}$ correspond to the INs ``Pizza delivery'', ``Cambridge (US)'' and ``Cambridge (UK)'' respectively. Then the IN for ``Pizza delivery in Cambridge (UK)'' would be represented by a superposition of $\varphi_p$ and $\varphi_{uk}$ vectors, as it is about both Pizza and Cambridge (UK). However the IN ``Cambridge'' represents classical ambiguity regarding the country, and thus it is represented as a mixture of pure IN vectors $\varphi_{uk}$ and $\varphi_{us}$. Thus a query ``Pizza delivery in Cambridge'' will be a mixture of superpositions.

This approach is later extended from a single Hilbert space of IN to multiple Hilbert spaces in \cite{Piwowarski:2010:QTB:1871437.1871450}. User IN is considered to be composed of several ``aspects'', which need to be addressed by the relevant documents. Each aspect is represented in a separate Hilbert space made up of IN aspect vectors for the aspect. For example, consider a query ``What tropical storms (hurricanes and typhoons) have caused significant property damage and loss of life?''. It comprises two IN aspects: ``tropical storms'' and ``significant damage/loss of life''~\cite{Piwowarski:2010:QTB:1871437.1871450}. The vectors for ``hurricane'' and ``typhoons'' are the IN aspect vectors for the tropical storm aspect of the query. As different aspect vectors belong to separate Hilbert spaces, the composite system corresponding to all IN aspects for the query is:
\vspace{-2mm}
\begin{equation}
\varphi_q = \varphi_1 \otimes \varphi_2
\end{equation}
where $\varphi_1$ and $\varphi_2$ are constructed in the same way as Equation \ref{query-density}. The probability of relevance of a document defined by the subspace $S_d$ in each Hilbert space would be $p(\otimes S_d|\varphi_q) = p(S_d|\varphi_1) \times p(S_d|\varphi_2)$.

The query representations for the above two approaches consider uniform weights to terms in mixtures and superpositions. However, the case of compound terms is not considered in \cite{Piwowarski:2010:QTB:1871437.1871450}. The issue is dealt with in \cite{queryalgebra}, which provides a sophisticated representation of query density matrices. The paper introduces a query algebra, which can be ``used to express relationship between query terms, thus allowing for more complex representations''~\cite{queryalgebra}. Several natural language processing (NLP) techniques such as Chunking and Dependency Parsing, are involved to identify different IN aspects and to characterize the relationship among terms within each aspect.
\\
\\
\noindent\textit{Polyrepresentation} 
\\
The concept of representing information systems as composite systems in separate Hilbert spaces is explored in \cite{Frommholz:2010:SPQ:1840784.1840802} for polyrepresentation of documents. A document may have different representations, based on different information sources, for example, text, author profiles, reviews and rating, etc. Each representation can correspond to different aspects of the information need of the user. Assume we have two Hilbert spaces representing a collection of books, one representing the authors and another for reviews. We have two authors $\ket{Smith}$ and $\ket{Jones}$ and two types of reviews $\ket{Good}$ and $\ket{Bad}$. Then a composite system of the two Hilbert spaces will be: 
\vspace{-2mm}
\begin{align}
(\ket{Smith} + \ket{Jones}) \otimes (\ket{Good} + \ket{Bad}) &= \\ \nonumber 
\ket{Smith}\ket{Good} + \ket{Smith}\ket{Bad} &+
\ket{Jones}\ket{Good} + \ket{Jones}\ket{Bad}
\end{align}
where the user is uncertain whether to read a book by James or Smith and also unaware of their ratings. However, an interesting case is that of non-separable states, where a user wants a book by Smith which is rated good or wants a book by Jones which is rated as bad. The composite system of user's IN in this case is given by a non-separable state:
\vspace{-2mm}
\begin{align}
\ket{Smith}\ket{Good} + \ket{Jones}\ket{Bad}
\end{align}
which reduces the uncertainty from the system point of view.

\subsubsection{Quantum inspired Language Models and Applications}

The Quantum Language Model (QLM) proposed by \cite{Sordoni:2013:MTD:2484028.2484098} combines the Vector Space Model and Probabilistic Language Model of classical IR via the Hilbert space formalism. The Quantum generalization of probabilities comes in the form of representing compound terms in queries and documents as superposition events, which have no classical analogue. This generalized Quantum probability model reduces to classical in case of using single terms only. More recently, a number of extensions of QLM have been made. 
\\
\\
\noindent\textit{Basic QLM}
\\
In QLM proposed in~\cite{Sordoni:2013:MTD:2484028.2484098}, a document or query is represented as a sequence of projectors. A projector represents a single term or compound term from the document/query. A document $d$ containing words from a vocabulary of size $N$ is represented as:
\vspace{-4mm}
\begin{align}
P_d = \{\pi_i : i = 1, ..., M\} \hspace{0.2cm} where \hspace{0.2cm} M \leq N
\end{align}
The Hilbert space is a term space of dimensionality $N$, where each vector $\ket{v_s}$ represents a term from the vocabulary. Thus the projector for a single term is $\pi_w = \ket{v_s}\bra{v_s}$. The vector for a compound term $\ket{v_{s_1..s_k}}$ is the superposition of all the vectors corresponding to the single terms:
\vspace{-4mm}
\begin{align}
\ket{v_{s_1..s_k}} = \sum_{i=1}^k{\sigma_i \ket{v_{s_i}}}
\end{align}
where $\sigma_i$ quantifies how much the compound term represents the single term $s_i$, and $\sum_{i=1}^k{|\sigma_i|^2}=1$. Thus in the same subspace, the representation of new term is created. This is not possible in traditional Vector Space Models because for every new term, single or compound, one has to add a new dimension to the vector space. Representing compound terms as superposition events solves that problem. Also, the compound term and the single terms in it are not disjoint and are related by the $\sigma_is$. 

Practically, in order to construct the projectors for a document, the terms co-occurring in the document within a fixed window of size $L$ are considered as compound terms. A language model is essentially a density matrix $\rho$, and for a document it is represented by projectors $P_d = \{\pi_1, \pi_2, ..., \pi_M\}$. It is obtained by maximizing the following function:
\vspace{-2mm}
\begin{equation}
L_{P_d}(\rho) = \prod_{i=1}^{M}{tr(\rho \pi_i})
\end{equation}
The language model is estimated using a generalization of an Expectation-Maximization based algorithm, called the $R\rho R$ algorithm~\cite{homodyne}. 

The language model for a query $\rho_q$ can be estimated in a similar way. The relevance of a document for a query can be calculated using a generalization of the KL divergence method called \textit{quantum relative entropy} or \textit{Von-Neumann(VN) divergence}~\cite{umegaki}. Given two language models $\rho_q$ and $\rho_d$, the scoring function is:
\vspace{-2mm}
\begin{align}
\Delta_{VN}(\rho_q || \rho_d) = -tr(\rho_q \log \rho_d)
\end{align}
where $tr(x)$ denotes the trace of the matrix $x$. Experimentally, the QLM has been shown to outperform a baseline language model and a Markov random fields (MRF) based model~\cite{mrf-model-baseline} (which was state-of-the-art at the time of publication of ~\cite{Sordoni:2013:MTD:2484028.2484098}) for document ranking.
\\
\\
\noindent\textit{Extended QLMs}
\\
Several extensions have been made to the basic QLM described above. \cite{Xie2015ModelingQE} propose to augment the QLM by making use of ``entangled'' terms. Based on the relation between Unconditional Pure Dependence (UPD) and Quantum entanglement~\cite{Hou2009CharacterizingPH, Hou:2013:MPH:2493175.2493177}, the UPD patterns are extracted from queries and documents, and the corresponding projectors are constructed. Instead of considering arbitrary compound terms, these UPD patterns are used as they show a statistical relationship similar to entangled systems.

Moreover, the $R\rho R$ algorithm used in QLM has a disadvantage in that it does not always converge. Hence a new global convergence algorithm is used in \cite{unsupervised_sentiment_analysis} for a Global Quantum Language Model (GQLM) but applied on twitter sentiment analysis tasks. Two dictionaries of positive and negative sentiment words are constructed. The global convergence algorithm constructs density matrices for the dictionaries and documents. Then using the quantum relative entropy, a document is projected onto each dictionary to determine its sentiment class. 

A Quantum language model based Query expansion approach is presented in \cite{Li:2018:QLM:3234944.3234970}. Using the GQLM described earlier, the language models for documents and query are constructed. The initial ranking is achieved using the Quantum relative entropy. Then a density matrix is constructed for the top $k$ retrieved documents. The top $n$ non-query terms, corresponding to the top $n$ diagonal elements of the density matrix, are selected as expanded terms. The top $n$ diagonal elements are in the order of the Quantum probabilities of the terms. Hence the advantage of using Quantum probabilities can be intuitively understood from here - the quantum probability reflects both the single term occurrence and the co-occurrence between terms. Hence, ``a term with a high frequency but a low co-occurrence with other terms may as well have a lower quantum probability than a terms with lower frequency but higher co-occurrence''~\cite{Li:2018:QLM:3234944.3234970}. After having formed the expanded query, a new GQLM is constructed for the expanded query and the documents are re-ranked accordingly. Experiments on the TREC 2013 and 2014 session track datasets show a better performance than the original QLM and another quantum model proposed in~\cite{quantum_weak_measurement}, which is based on user interactions. Indeed, there have been various extensions of QLM that adapt to user interactions~\cite{QiuchiLi2015ModelingMR, Li2016, two_state_vector}, which will be discussed in Section 3.2.

The QLM is also extended within a neural network structure in \cite{Zhang2018EndtoEndQL} for Question Answering (QA), while the authors mention that the model can also be applied to other IR tasks, such as ad-hoc retrieval. Using word embeddings as vectors, a density matrix for each sentence is constructed, for both questions (as queries) and answers (as documents). The density matrix represents a mixture of semantic spaces. A joint representation of queries and documents is constructed by multiplying the density matrices for queries and documents. Then a convolution layer is applied over this joint representation followed by pooling, a fully connected layer and a softmax layer. The binary output of the softmax layer represents probabilities of relevance and non-relevance of the answer with respect to the question. This process is repeated for each question and answer pair and a ranking based on their relevance probabilities is produced. This model achieved MAP and MRR scores of 0.7589 and 0.8254 on the TREC-QA dataset, which was 2.46\% and 3.24\% improvement over a neural model TANDA \cite{trec-qa-baseline}, which was state-of-art at the time when the paper was published\footnote{Note that the current state-of-the-art BERT-based neural model for TREC-QA has achieved MAP and MRR of 0.943 and 0.974 respectively~\cite{trec-qa-state-of-art}.}.

A Quantum many body wave function based language model is presented in~\cite{Zhang_many_body}. The aim is to create a language model which addresses the challenges in word combinations, where each of the individual words can possess multiple meanings. Different meanings of a word are represented as different basis vectors of a Hilbert space. The state vector for a word is a superposition of different base vectors corresponding to different meanings of the word. The state vector of sentence is represented as a tensor product of the individual word's state vectors. This is termed as a local representation, and a similar global representation of the language model is constructed using another corpora, to account for unseen words and unseen compound words. The global representation is projected onto the local representation akin to the smoothing process in classical language models. As the global representation is a higher rank tensor, it is decomposed using tensor decomposition techniques. The projection of the reduced tensor onto the local representation tensor is realized in the form of a convolutional neural network. While it is unable to outperform the above mentioned model \cite{Zhang2018EndtoEndQL} on the TREC-QA dataset, it performs significantly better on the WikiQA dataset\footnote{The TANDA model mentioned above currently gives the best performance on WikiQA dataset (MAP and MRR of 0.92 and 0.933 respectively, as compared to 0.695 and 0.71 by~\cite{Zhang_many_body}).}. 
\\
\\
\noindent\textit{Other Quantum-inspired language models}

In \cite{blacoe_quantum_distributional}, the Quantum theoretic framework is used to construct a syntax-aware semantic model. It also takes word order into account, unlike the QLM. Firstly, for each sentence, dependency parsing is performed and a set of dependency relations are extracted. This set is partitioned into clusters of syntactically similar relations, and each cluster is assigned a Hilbert space. Each Hilbert space has the word vectors as the basis vectors. The state vector in each Hilbert space is a superposition of the word vectors, which are dependencies of the same word. The state vector for a given word is written as a tensor product of the state vectors in all the Hilbert spaces. A complex phrase is ascribed to the state vector. A word occurring in different senses will have different state vectors, which are then superposed to get the final vector for each word.  It is then converted into a density matrix, and the density matrices of the occurrences of the word in different documents are added up. The similarity between two words can be measured using the trace rule, which essentially takes the pair-wise inner products of the state vectors. This allows the words to "select" each other's context and should lead to more accurate similarity values. Experiments done on word similarity and word association tasks reveal a better performance than some classical models. This method is extended in \cite{BlacoeQI} to construct density matrices for sentences. To create a density matrix for a sentence, first the dependency parsing tree is constructed. For each node in the tree, its dependencies are projected onto it and the post projection states are summed up together with the density matrix of the node. This procedure is performed recursively until the whole sentence is covered. The method is tested on the paraphrase detection task with the Microsoft Paraphrase Detection dataset, and shows better accuracy and F1 scores than a recent neural network model reported in \cite{msprc-baseline-blacoe-qlm}.

In~\cite{towards_QLM}, an n-gram language model inspired from QT is introduced, with application to speech recognition. Unlike the quantum inspired language models presented earlier, this paper makes use of the unitary evolution of a quantum state in time, e.g. $\rho_{t+1} = U\rho_t U^{\dagger} $, where $U$ is a unitary operator which changes the state of a system $\rho_t$. To measure the probability of a word $w$, the system state is projected onto the state of the word $w$ using the projector $\Pi_w = \ket{w}\bra{w}$. Probabilistic information about a sequence of words $w_1, w_2, ..., w_n$ is encoded in a density matrix built using the following process:
\vspace{-3.5mm}
\begin{align}
    p(w_1; \rho_0, U) &= tr(\rho_0, \Pi_{w_1}) \nonumber \\
    \rho_1^{'} &= \frac{\Pi_{w_1}\rho_0\Pi_{w_1}}{tr(\Pi_{w_1}\rho_0\Pi_{w_1})} \\ \nonumber
    \rho_1 &= U\rho_1^{'}U^{\dagger}
\end{align}

where $\rho_0$ is the initial state of the system, and $\rho_1$ is state of the system after processing the first word in the sequence. The unitary matrix $U$ is responsible for the time evolution of the system and $tr(x)$ stands for the trace operation. The final probability of the whole sequence becomes:
\vspace{-2mm}
\begin{equation}
    p(w_i|w_1, ..., w_{i-1}) = tr(\rho_{i-1}\Pi_{w_i})
\end{equation}

One possible issue arises here. Continuously projecting and collapsing the system state to individual words may remove any quantum effects from the system, i.e. the system reduces to a classical markov model like system. To address the issue, the system is coupled with an ancillary system to avoid the complete collapse. A D-dimensional Hilbert space represents the ancillary system and thus the composite system has the Hilbert space $H_2 = H_{ancilla} \otimes H$. The new projectors for the composite space are given by $\Pi_w^{(2)} = I_D \otimes \Pi_w$. The advantage of doing this is that the time evolution of the composite system can give rise to non-trivial correlations between them (analogous to non-separability and entanglement) so that even when the state of the word sequence is collapsed, some information is retained in the ancillary part (owing to their non-trivial correlations). The words are represented in low-dimensional vectors and for each dimension, a unitary matrix is assigned for the composite system. The parameters are learned by minimizing the perplexity of the corpus of sentences. The perplexity is given by:
\vspace{-1mm}
\begin{equation}
    \Gamma(\rho_0, U) = exp(-\frac{1}{C}\sum_{w \in S}{\log p(w|\rho_0, U)} 
\end{equation}

Experiments on the TIMIT dataset show that this n-gram quantum language model has a lower perplexity than the state-of-the-art deep neural network architectures like RNN and RNN-LSTM. Although the paper reports an application of the proposed language model in speech recognition, it would be interesting to use it to construct document and query language models.

\subsubsection{Quantum-inspired Ranking}

The research on quantum-inspired ranking has been done from two perspectives: quantum probability ranking principle and quantum-like measurement.
\\
\\
\textit{Quantum Probability Ranking Principle}
\\
According to the Probability Ranking Principle ~\cite{prp_robertson}, an IR system should rank the documents for a user IN in a decreasing order of their probability of relevance. It makes the assumption that ``the relevance of a document to an information need does not depend on other documents''~\cite{Zuccon2009}. However, in real world situations, judgment of documents by a user is influenced by its previously judged documents~\cite{Eisenberg1988}. ``The utility of a document may become void if the user has already obtained the same information''~\cite{Zuccon2009}. This 'interference' between documents can be due to information overlap between documents or a change in the IN, and is accounted for in a Quantum Probability Ranking Principle (QPRP)~\cite{Zuccon2009}. QPRP draws an analogy~\cite{Melucci2010} with the Double Slit Experiment by assuming the two slits to be two documents $A$ and $B$ which the user judges for a query. The position $x$ on the screen corresponds to the event that the user is satisfied by the documents $A$ and $B$, and decides to stop the search. If $A$ is first document presented to the user, we have $p_{AB}(x)$ as the probability that the user stops the search at document $B$. In the Double slit experiment, if slit $A$ is fixed and slit $B$ is varied in dimensions, which is analogous to having different documents listed after document $A$, we get $p_{AB_i}(x)$ as ``the probability of stopping the search process after seeing documents $A$ and $B_i$''~\cite{Zuccon2009}. The problem then boils down to finding which configuration of slits (set of documents) $AB_i$ exhibits maximum value of $p_{AB_i}(x)$. 

For the classical case, if there is no interference, i.e. only one of the $B_i$ slit is opened at a time, we have "$p_{AB_i}(x) = p_A(x) + p_{B_i}(x)$"~\cite{Zuccon2009}:
\vspace{-2mm}
\begin{align}
\arg\max_x(p_{AB_i}(x)) = \arg\max_x(p_A(x) + p_{B_i}(x)) = \arg\max_x(p_{B_i}(x))
\end{align}
However, in the quantum case, with all slits open, or all documents considered by the user till $B_i$, $p_{AB_i}(x) = p_A(x) + p_{B_i}(x) + I_{AB_i}(x)$, where $I_{AB_i}(x)$ is the interference term. Thus:
\vspace{-2mm}
\begin{align}
\arg\max_x(p_{AB_i}(x)) &= \arg\max_x(p_A(x)  + p_{B_i}(x) + I_{AB_i}(x)) \\ \nonumber 
&= \arg\max_x(p_B(x) + I_{AB_i}(x))
\end{align}
Hence the best choice of document to rank after $A$ is not the one whose relevance probability is maximum, but rather the one whose sum of individual relevance probability and the interference term with $A$ is maximum. Hence, between two documents $B$ and $C$, $B$ is ranked before $C$ if and only if:
\vspace{-3mm}
\begin{equation}
p_B(x) + I_{AB} \geq p_C(x) + I_{AC}(x)
\end{equation}

Recall from Equation \ref{interference-term} that the interference term depends upon the phase difference of the probability amplitudes of two quantum systems. Thus:
\vspace{-4mm}
\begin{align}
p_{AB}(x) = p_A(x) + p_B(x) + 2\sqrt{p_A(x)}\sqrt{p_B(x)}\cos(\theta_{AB})
\end{align}
In~\cite{Zuccon2009}, the authors did not give details of how to estimate the interference term. This estimation is done in an application of the QPRP to subtopic retrieval~\cite{Zuccon:2010:UQP:2128344.2128384}. Subtropic retrieval is a task of providing a list of documents which covers all possible topics (IN aspects) relevant to the user IN. It advocates a more diverse ranking of documents, to achieve a minimal redundancy. Thus redundant documents can be assumed to be destructively interfering (negative interference term) and the documents having exclusive information be positively interfering. The $\cos(\theta)$ part of the interference term is estimated as the Pearson's correlation between the term vectors of two documents. The term vectors are constructed using the BM25 scheme. Experiments show that the QPRP based ranking for subtopic retrieval performs better than a model based on Portfolio Theory\cite{wang2009portfolio} (then state-of-the-art) for subtopic retrieval on the ClueWeb09-B collection.
\\
\\
\noindent\textit{Quantum Measurement inspired Ranking }
\\
Another method for document ranking using Quantum probabilities is discussed in \cite{Zhao2011}, called Quantum Measurement inspired Ranking (QMR). Document retrieval process is considered to be similar to a photon polarization process. A photon has a Horizontal or Vertical polarization, which can be measured by a polarizer. There also exist superposition states of both vertical and horizontal polarizations, which is detected by a horizontal or vertical polarizer rotated at a $45$ degree angle. 
\vspace{-6mm}
\begin{align}
\ket{\nwarrow} = \frac{1}{\sqrt{2}}(\ket{\uparrow} + \ket{\downarrow})
\end{align}
Superposition states can be generated by passing a horizontal or vertically polarized photon through the rotated polarizer. Mathematically, the vertical and horizontal polarizers form an orthonormal basis of a two dimensional Hilbert space. The rotated polarization state forms another orthonormal basis in the same Hilbert space. In the analogy, the first round of document retrieval for a query is analogous to the measurement along the vertical or horizontal basis. Then, a second round retrieval is performed to re-rank the documents by comparing all retrieved documents with the top $k$ documents. This is analogous to passing the photons coming from a horizontal or vertical polarizer through the rotated polarizer.  Mathematically this is formulated as projecting a vector represented in one basis onto the subspace generated by another rotated basis. 

In the first round of retrieval, let $\ket{\uparrow}$ and $\ket{\downarrow}$ denote relevance and non-relevance of document respectively for a query. Then a document $d$ with relevance probability $|\alpha_d|^2$ is represented in the first round as:
\vspace{-2mm}
\begin{align}
\ket{d} = \alpha_d\ket{\uparrow} + \beta_d\ket{\downarrow}
\end{align}
Taking the simple case of $k=1$, let the topmost document in the first round of retrieval be represented as: 
\vspace{-2mm}
\begin{align}
\ket{t} = \alpha_t\ket{\uparrow} + \beta_t\ket{\downarrow}
\end{align}
Then, re-ranking is done by representing the document $d$ in terms of $t$:
\vspace{-2mm}
\begin{align}
\ket{d} = \lambda\ket{t} + \mu\ket{\widetilde{t}}
\end{align}
where $\lambda = \alpha_d\alpha_t + \beta_d\beta_t$ (see appendix in ~\cite{Uprety:2018:IOE:3234944.3234972} for a proof). The probability of relevance of the document $d$ when re-ranking is performed using the top-ranked document of first round is the square of the projection of $d$ onto $t$, which is $|\lambda|^2$, multiplied by the probability of relevance of $t$, which is $|\alpha_t|^2$:
\vspace{-2mm}
\begin{align}
p(d|t) =  |\lambda\alpha_t|^2
\end{align}
When $d = t$, then $\lambda = 1$ and the probability becomes $|\alpha_t|^2$, the original probability of relevance of the top-ranked document. The QMR approach performs significantly better than the QPRP on four TREC collections - WSJ9872, AP8889, ROBUST04 and WT10G on MAP ranking metric.

Quantum-inspired ranking has also been used to solve the query drift problem, which is defined as the inferiority of results obtained on query expansion, than the original query. This is because the underlying intent of the query might change upon expansion. Several solutions have been proposed for the query drift problem using pseudo relevance feedback~\cite{Zighelnic:2008:QPR:1390334.1390524}, focusing on the combination of document scores in the ranked lists of documents based on the original query and the expanded query. For example:  
\vspace{-1.5mm}
\begin{itemize}
\item CombMNZ rewards documents that are ranked higher in both original retrieval list and second retrieval list by adding the relative score of a document in each of the two lists.
\end{itemize}

\begin{itemize}
\item Interpolation technique makes a weighted addition of relative scores in the two lists.
\end{itemize}

%\begin{itemize}
%\item The re-rank method ranks the pseudo relevance feedback %based documents based on their original scores.
%\end{itemize}

In \cite{Zhang2011InvestigatingQP}, a document $d$ is represented in terms of relevance and non-relevance for a query $q$:
\vspace{-2mm}
\begin{align}
\ket{d} =  a_d\ket{q} + b_d\ket{\widetilde{q}}
\end{align}
where $\ket{q}$ and $\ket{\widetilde{q}}$ represent the vectors for relevance and non-relevance of $d$ with respect to $q$, respectively. In terms of the expanded query $q^e$, the document is represented as:
\vspace{-2mm}
\begin{align}
\ket{d^e} =  a_d^e\ket{q^e} + b_d^e\ket{\widetilde{q^e}}
\end{align}
``To prevent query drift, the existing fusion models in \cite{Zighelnic:2008:QPR:1390334.1390524} directly combine the two probabilities $|a_d|^2$ and $|a_d^e|^2$''~\cite{Zhang2011InvestigatingQP}. The CombMNZ reduces to: 
\vspace{-2mm}
\begin{align}
(\delta_q(d) + \delta_q^e(d)).(\delta_q(d)|a_d|^2 +  \delta_q^e(d)|a_d^e|^2)
\end{align}
where $\delta_q(d) = 1$ if $d$ is relevant to query $q$. Similarly, the interpolation method becomes:
\vspace{-2mm}
\begin{align}
\lambda\delta_q(d)|a_d|^2 + (1 - \lambda)\delta_q^e(d)|a_d^e|^2 \hspace{0.3cm} 0\leq \lambda\leq 1
\end{align}
However, the two probabilities $|a_d|^2$ and $|a_d^e|^2$ are under different basis and we need to write one in terms of the other. The Quantum Fusion Model (QFM) proposed in \cite{Zhang2011InvestigatingQP} does that and the final outcome combines the probabilities in the following way:
\vspace{-4mm}
\begin{align}
(\delta_q(d)|a_d|^2).(\delta_q^e(d)|a_d^e|^2)
\end{align}
Thus the Quantum based model is a multiplicative model, while the classical models are additive. Another slightly modified version is:
\vspace{-4mm}
\begin{align}
(\delta_q(d)|a_d|^2).(\delta_q^e(d)|a_d^e|^2)^{1/\eta}
\end{align}
where ``a small $\eta$ can make scores of different documents retrieved for $q^e$ more separated from each other, leading to more distinctive scores''~\cite{Zhang2011InvestigatingQP}. The QFM achieves a better performance than the CombMNZ and interpolation methods in terms of Mean Average Precision (MAP) of retrieved documents.

\subsubsection{Multimodal Information Retrieval}

Despite the wide application of QT in text-based IR, limited attention has been paid to multimodal IR, which is of increasing significance in many applications. The work in this area can be divided into feature level fusion and decision level modality fusion.
\\
\\
\noindent\textit{Feature Level Fusion}
\\ 
 Initially, Wang et al. \cite{wang2010tensor} exploited tensor product of Hilbert spaces to fuse textual and image features for circumventing the heuristic combination of uni-modal feature spaces. In particular, textual and visual features are combined in a similar way as non-separable states of a Quantum system. The authors claim that the proposed modality fusion approach is able to capture cross-modal dynamics, i.e., interactions across different modalities (e.g., text and image modalities). In each single modality feature space, documents are formulated as a superposition of terms. These terms are words from a vocabulary for the text representation and visual words for the image representation. For instance, in the
textual feature Hilbert space denoted as $H_T$: 
\vspace{-2mm}
\begin{align}
\ket{d_T} = \sum_{i}{w_{t_i} \ket{t_i}},
\end{align}
where the squared amplitude $w_{t_i}^2$ equals the probability of a document to be about the term $t_i$ with $ \sum_{i}{w^2_{t_i}}=1$. Similarly, for the image modality the formulation is as follows:
\vspace{-2mm}
\begin{align}
\ket{d_V} = \sum_{i}{w_{v_i} \ket{v_i}},
\end{align}
where $v_i$ represents visual words in the image Hilbert space $H_V$. Each pure state is modelled through a  density matrix. In mathematical language, each density matrix is defined as the outer product of a superposition state. For example, for the textual representation, the density matrix is:
\vspace{-2mm}
\begin{align}
\rho_{d_T} = \sum_{i} {p_i \ket{d_{T_i}}\ \bra{d_{T_i}} },
\end{align}
where  $p_i$ is the probability of the state being in the basis state $\ket{d_{T_i}}$.
Then, the text and image modalities are fused by taking the tensor product of the text and image Hilbert spaces as follows:
\vspace{-2mm}
\begin{align}
\rho_{d_{TV}} = \rho_{d_T} \otimes \rho_{d_V} + \rho_{correlation},
\end{align}
where $\rho_{d_T} and \rho_{d_V}$ are the textual and visual density matrices respectively, and $\rho_{correlation}$ is the density matrix capturing cross-modal interactions between the text and image features. The resultant product is still a valid density matrix. Finally, for measuring the similarity between a multimodal document and query, the trace rule is used as follows: 
\vspace{-2mm}
\begin{align}
   sim(d,q) = trace(\rho_{d_{TV}} \cdot \rho_q),
\end{align}
where $\rho_{d_{TV}}$ and $\rho_q$ are multimodal document and query density matrix representations respectively. %The result of the measurement is a scale real number.

For capturing cross-modal interactions across the two modalities, two statistical approaches were explored: (a) the maximum feature likelihood that associates text with the maximal likely image features; and (b) the mutual information matrix that measures the mutual dependence between the two modalities. Experiments show that even without considering the correlation between text and image features, the pure tensor product approach outperforms other methods such as the use of image features or text features individually, or the concatenation of text and image features. However, such a method suffers from exponentially increasing computational complexity, as the outer product over multiple modalities results in high dimensional tensor representations. The experiments also show that the two proposed statistical methods are trivial to capture cross-modal interactions. For example, simple visual features were used, such as colour histograms, which can hardly be associated with high-level semantics.  A more robust statistical approach, such as the cross-modal factor analysis \cite{atrey2010multimodal}, might be more effective. Another issue was that images with limited or no annotation were lowly ranked or not retrieved at all. This implies that tensor product cannot manipulate missing values, which becomes a common problem in a real-world scenario. An automatic annotation task might circumvent the above problem. Also, assuming orthogonality of dimensions disregards any semantic overlap. This was an issue for textual space as the dimensions representing words need to represent language attributes such as polysemy and synonymy. Nowadays, we can address such issue by exploiting neural network language technologies \cite{pennington2014glove,devlin2019bert} for constructing text vector spaces with compact semantic information.

Kaliciak et al. \cite{kaliciak2011contextual} followed up with the previous model, aiming to solve the problem of missing modalities, e.g., when images are not annotated. They proposed two approaches to alleviate this problem, which can be easily integrated with the tensor-based fusion method. The first approach projects an un-annotated image onto a subspace generated by subsets of annotated images. In particular, by exploiting the Born rule, the square projection on the basis states results in a probability distribution of terms for each un-annotated image. The second approach alternatively utilizes the trace rule to calculate the similarity between an annotated and un-annotated image. Images are formulated as density matrices, entailing a probability distribution of terms.  The results showed that such approaches under-performed the standard clustering techniques. The result might be related to the assumption that ``the correlation at the image-level (i.e., images referring to the same topic) are stronger than the correlations based on the proximity between image terms (i.e., instances of image words)''~\cite{kaliciak2011contextual}.

Later on, Kaliciak et al. \cite{kaliciak2013combining} proposed a quantum-inspired framework for a cross-modal retrieval task. That is, given a text query, to retrieve the most relevant images. They first constructed a common Hilbert space by taking the tensor product of image and text density matrices. Both text queries and image documents are represented in the joint Hilbert space. In this joint space, they also utilized a mechanism of trans-media pseudo-relevance for re-ranking retrieved images. Then, a projection measurement measures the relevance between the text query and each image document represented on the same space.
\\
\\
\noindent\textit{Decision-level Fusion}
\\
Decision level fusion combines uni-modal classification results to reach a final decision. Despite being a common approach for fusing different modalities, only preliminary studies have investigated quantum-inspired decision level approaches for IR tasks. Gkoumas et al. \cite{dimitris_QI} investigated non-classical correlations between mono-modal decisions on a pair of text-image documents for a multi-modal retrieval task. In principle, non-classical correlations or quantum correlations are stronger than classical correlations due to latent contextual influences. In that study, the authors investigated the existence of this kind of non-classical correlation through the Bell inequality (CHSH inequality) violation in a small-scale experiment on the ImageCLEF dataset. Although they did not find a violation of the CHSH inequality, the experiment design provides useful insights for future investigations into such non-classical contextual correlations.

Quantum-inspired modality fusion models have also been developed for multimodal sentiment analysis. Sentiment is one of the factors considered by users in judging certain types of documents (e.g. news articles, blogs). Sentiment label can be considered as a feature in predicting relevance~\cite{Fuhr_info_nutrition}.

Zhang et al.~\cite{zhang2018quantum} proposed a quantum-inspired decision level modality fusion approach for image-text sentiment analysis. Even this task is far from IR tasks, the approach could be fruitful for IR tasks as well. In particular, both text and image information is associated with density matrices which use the same globally convergent algorithm mentioned earlier in the case of extended QLMs to estimate the density matrices. In this way, density matrices capture the cross-modal interactions. Additionally, the human cognitive interference phenomenon caused when a user is exposed to conflicting text and image information channels, is also considered as analogous to quantum interference. Though, the interference term is treated as a single parameter and adjusted experimentally. The results suggest that, when the Cosine of the interference term equals 0.3, the model achieves the best performance. Moreover, the accuracy is the highest when a user pays more attention to the text instead of image modality, assigning weights 0.7 and 0.3 for the text and visual representation respectively. When the weights are reversed, the model performs the lowest. This is an interesting outcome since it helps us understand under which conditions the quantum-like interference works at the decision level and enhances explainability over the modality fusion process. Overall, large-scale experiments show that the proposed approach outperforms a wide range of state-of-the-art baselines. 

A combination of Long Short Term Memory (LSTM) and a quantum-inspired framework for conversational sentiment analysis was proposed in \cite{zhang2019quantum}. In particular, words are represented as pure states in a real-valued Hilbert space. Then, a sentence is formulated as a mixture density matrix of pure states, i.e., unit vectors, which further is processed by a CNN, resulting in a dense representation. Next, the output of CNN is fed into an LSTM cell to make a  decision. Considering conversation sentiment analysis contains time-series and thus requiring fusing time-varying signals, the authors exploited a sequence of LSTM cells and the concept of quantum-inspired measurement, namely weak measurement, to model inter-speaker sentiment influences over a dialogue.

\cite{zhang2020quantum} is a follow up of \cite{zhang2019quantum} by extending the framework with two modalities, namely, text and visual modalities. Specifically, each modality is represented in an individual real-valued Hilbert space. The exact pipeline with \cite{zhang2019quantum} is followed to predict unimodal sentiment judgments. Then, the concept of quantum interference has been exploited to fuse text and visual sentiment judgements. Comprehensive experiments on two benchmarking datasets for conversational human language analysis showed that the proposed quantum-inspired framework beats the state-of-the-art performance for the video emotion recognition task. It is to be noted that conversational sentiment analysis is an important feature in conversational IR tasks.

\subsubsection{Quantum inspired Representation Learning}

The quantum-inspired representation and ranking models depend on the construction and learning of Hilbert spaces. These are developed or applied in areas like lexical semantic spaces, topic modelling, word embeddings, and text classification.
\\ 
\\
\\
\noindent\textit{Lexical Semantic Spaces}
\\
The first connections between QT and semantic spaces were established in \cite{aerts_czachor}. In \cite{BruzaC05}, one such connection is presented using the Hyperspace Analogue to Language (HAL) model~\cite{Lund1996, Burgess1998HAL}. For a vocabulary of $N$ words, the HAL algorithm constructs an $N \times N$ matrix by sliding a window of length $l$ over a text corpus, thus capturing word co-occurrences within the window. Each element of the matrix measures word co-occurrence and in one way, word similarity.  Each window is considered as a semantic space and approximates the context or the sense associated with the word. The semantic space for a word is computed in terms of the sum of semantic spaces. If there are $y$ windows around the word $w$ and $x$ of them deal with a particular context $i$, then the semantic space $S_i$ occurs with probability $p_i = \frac{x}{y}$ and the semantic space for word $w$ can be written in terms of context semantic spaces as:
\vspace{-3mm}
\begin{equation}
S_w = \sum_{i=1}^{m}{p_iS_i}
\end{equation}
This formula is the same as that of a mixed density matrix written as a mixture of density matrices of pure states. Thus the context of words can be considered as pure states. HAL is also used in \cite{Hou2009CharacterizingPH,Hou:2013:MPH:2493175.2493177} to model word correlations like Quantum correlations of non-separable states.

The explicit term occurrence based approaches are insufficient to capture hidden semantics. The advent of machine learning techniques have opened up a door to learning semantic spaces based on topic modelling and word embedding. 
\\ 
\\
\noindent\textit{Interference Topic Model}
\\
The analogy to Quantum interference is used in \cite{Sordoni:2013:MLT:2505515.2507854} for modeling interactions between topics. Topic modeling is used to discover hidden themes in text collections. A topic is defined as a probability distribution over a vocabulary and a document is a mixture of one or more such topics. Finally, every word in a document is supposed to come from one of the topics. The probability of a term $w$ in a document model $\theta_d$ with $k$ topics is given as~\cite{Sordoni:2013:MLT:2505515.2507854}:
\vspace{-4mm}
\begin{align} \label{classical-topic}
p(w|\theta_d) = \sum_{k}{p(w|z=k,\phi)}p(z=k|\theta_d) = \sum_{k}{\theta_{dk} \phi_{kw}}
\end{align}
where $w \in \{1,...,N\}$ denotes a word from the vocabulary. $z \in \{1,...,K\}$ is the index for a topic. $\theta_{d}$ denotes a document where $\theta_{d} = (\theta_{d1},...,\theta_{dk})$ , $\theta_{di}$ being the ``topic proportions for the document''~\cite{Sordoni:2013:MLT:2505515.2507854}. $\phi$ is a $N \times K$ matrix representing the distribution of topics over terms. 

Consider the case of two topics: `war' and `oil'. The term `Iraq' is present in both topics. Now if a document contains both topics, still the probability of term 'Iraq' in the document is less than the maximum of its probability in either of the topics. Mathematically speaking~\cite{Sordoni:2013:MLT:2505515.2507854},
\vspace{-2mm}
\begin{align}
p(w=Iraq|\theta_d) &= p(Iraq|war)*p(war|\theta_d) + p(Iraq|oil)p*(oil|\theta_d) \\ \nonumber
p(Iraq|\theta_d) &\leq \max(p(Iraq|war), p(Iraq|oil))
\end{align}
However, the probability of the term `Iraq' occurring in the document should be significantly more given it contains topic `war' and `oil'. Current topic models do not consider the interference or relation between two topics when generating a word. They assume the topics to be independent. To capture topic dependence via Quantum probabilities, \cite{Sordoni:2013:MLT:2505515.2507854} assumes a Hilbert space where each dimension corresponds to a word from the vocabulary. Then, each topic is a vector in this Hilbert space $z_k$, which is a superposition of vectors corresponding to the terms. Thus we have:
\vspace{-2mm}
\begin{equation}
\ket{z_k} = \sum_{w}z_{kw}\ket{e_w} = \sum_{w}{\sqrt{\phi_{kw}}}e^{i\varphi_{kw}}\ket{e_w}
\end{equation}
where $\sqrt{\phi_{kw}}e^{i\varphi_{kw}}$ is the complex amplitude for the topic $\ket{z_k}$ in state $\ket{e_w}$ and $|\sqrt{\phi_{kw}}e^{i\varphi_{kw}}|^2 = p(w|z=k,\phi)$. A document can be represented as a superposition of topic states, with the coefficients being the proportion of topic in the document. 
\vspace{-2mm}
\begin{equation}
\ket{d} = \frac{1}{Z_d}(\sum_{k}{\sqrt{\theta_{dk}}\ket{z_k}})
\end{equation}
where $Z_d$ is a normalization constant. The projection of a document vector onto a word vector is given as: 
\vspace{-2mm}
\begin{align}
d_w = \bra{e_w}\ket{d} \propto \sum_{k}{\sqrt{\theta_{dk}\phi_{kw}}e^{i\varphi_{kw}}}
\end{align}
The probability of a term in the document is given by:
\vspace{-2mm}
\begin{align}
p(e_w^+e_w) &= |\bra{e_w}\ket{d}|^2 \propto |\sum_{k}{\sqrt{\theta_{dk}\phi_{kw}}e^{i\varphi_{kw}}}|^2  \\ \nonumber
&= \sum_{k}{\theta_{dk}\phi_{kw}} + 2\sum_{i<j}{\sqrt{\theta_{di}\theta_{dj}}\sqrt{\phi_{iw}\phi_{jw}}\cos(\varphi_{iw} -  \varphi_{jw})}
\end{align}
This equation represents the proposed interference-topic model. The first part of the expression on the right hand side corresponds to the classical topic model given in Equation~\ref{classical-topic}, and the second is ``the interference term which boosts or penalizes the probability for term $w$ in the final document model depending on the phase differences $\varphi_{iw} -  \varphi_{jw}$''~\cite{Sordoni:2013:MLT:2505515.2507854}. For a particular word, if a pair of topics are in the same phase then, $\varphi_{iw} -  \varphi_{jw} = 0$ and $cos(\varphi_{iw} -  \varphi_{jw}) = 1$. This increases the probability of seeing the word $w$ in the document. For the phase difference of $\frac{\pi}{2}$, the interference term vanishes and the classical topic model is recovered. In their experiment, \cite{Sordoni:2013:MLT:2505515.2507854} estimated the interference term using a similarity measure between the topic distributions, such as the Cosine similarity. The topic model helps in relevance ranking in IR by providing a better match for queries and documents, beyond the term level. This Quantum-,inspired topic model is applied to retrieval tasks like the TREC newswire corpora and performs better than the classical topic model.
\\
\\
\noindent\textit{Complex Numbers}
\\
QT in its most general formulation uses complex numbers in its representations and computations. Without the use of complex amplitudes, for example, the interference effects will be restricted to only positive and negative interference values ($+1$ and $-1$), while not utilizing the full range of possibilities in between. Therefore, it is imperative that Quantum models outside of Physics which are directly or indirectly making use of the superposition and interference phenomena use complex numbers in representation of state vectors, in order to maximally exploit the power of quantum probabilities. However, it is difficult to get an intuition as to how to represent concepts, objects, terms, decisions, etc. using complex numbers.

\cite{Rijsbergen:2004:GIR:993731} proposed using inverse document frequency (idf) of a term as the imaginary part and the term frequency (tf) as the real part of a complex number. In \cite{10.1007/Guido_complex_num_ICTIR}, this proposal was investigated and found to be performing poorly than the baseline Vector Space models. In \cite{10.1007/complexIR_Wittek_Guido}, different types of word semantics are combined using a complex Hilbert Space. The main idea is to represent distributional semantics, like the word co-occurrence information as real part and represent ontological information about words as the imaginary part of a complex valued vector. The real vectors are constructed using the technique of Random Indexing, where a word vector is constructed using the vectors that represent the contexts of the word. A document is then represented as a sum of its word vectors. Besides, using the technique of concept indexing, a document is also represented as a Bag of Concepts vector. Each word is mapped to one or more concepts from a medical ontology. These two representations are merged into a single complex vector. Thus, similarity between two documents can be calculated as the inner product of the two complex vectors which will reflect both the distributional and ontological similarity. This model is used in the IR task of TREC Medical Records Track and the retrieval effectiveness is found to be better than either the term-based only or concept-based only approaches in terms of the Mean Average Precision (MAP) and Precision@10 metrics.
\\
\\
\noindent\textit{Quantum-inspired Neural Representation Models}
\\
In \cite{DBLP:conf/rep4nlp/LiUWS18}, the challenge of emergent meaning and sentiment of a combination of words is addressed. They hypothesize that humans do not associate a single meaning or sentiment to a word. A word contributes to the meaning or the sentiment of a sentence depending upon the other words it is combined with. For example, the words `Penguin' and `Flies' (Verb) might be neutral in polarity individually, but the phrase `Penguin Flies' is of negative polarity. Similar examples can be constructed for sentiment of sentences. This is compared to the Quantum interference phenomenon where two superposed quantum states interfere and the final outcome depends upon their relative phase. As such a word embedding model using complex numbers is introduced. Each word is represented by a complex vector. It comprises a real part that holds word co-occurrence information, and a complex phase that captures abstract combinatory information like the sentiment factor. A sentence is considered as a superposition of words and thus a sentence vector can be constructed as a density matrix. This density matrix is learnable from labelled data, along with a projection matrix, which is used to calculate the probabilities assigned by the sentence vector. For a sentiment classification task, the projection matrix is used to classify the sentiment of the sentence according to the trace rule of calculating probabilities. The proposed model outperformed some word-embedding based models.

Wang et al. \cite{wang2019semantic} proposed an end-to-end quantum-inspired neural framework for text classification. In particular, words are represented as pure states in a complex-valued Hilbert space, while sentences as mixture of pure states (i.e., words). Hence a sentence is represented in a mixture density matrix.  In this work, phases are not defined explicitly but learnt through a backpropagation algorithm. Having said that, the exploitation of complex values is pivotal to the formulation of quantum concepts. Then, a set of measures is applied to the mixture density matrix, resulting in a sequence of scalar values. Practically, the measurements are related to high semantic concepts. Finally, a softmax function normalises the output of measurements into a probability distribution before classifying the sentence. Comprehensive experiments on six different datasets demonstrated the effectiveness of the proposed method against some neural models.

This framework was extended in \cite{li2019cnm} for a question-answering task. The words are formulated as pure states in a complex-valued Hilbert space. Though, in contrast to \cite{wang2019semantic}, each word is represented in a pure density matrix while a sliding window is applied to the sentence, generating a local mixture density matrix for each local window. Both question and answer are represented by a sequence of mixture density matrices. Then, the same set of common measurements is applied to those density matrices in the sense that they share common semantic concepts. A max-pooling function is performed over the measurement output components, resulting in a dense representation before the matching process through the Cosine similarity measurement. The proposed complex-valued network for matching achieved comparable performances to strong CNN and RNN baselines on two benchmarking question answering (QA) datasets.

In~\cite{Zhang_many_body}, a quantum many-body wave function is used to model the semantic meaning of words within a local context, i.e., sentences, and a global context, i.e., corpus. In particular, each word is represented by different base states in the sense that each basis corresponds to a different word meaning. First, they model the word meaning within a sentence and corpus by the tensor product of basis states and then project the global tensor representation onto the local one. This results in a high dimensional tensor, capturing interactions of words within a sentence and corpus.  The high dimension resulted tensor representation is further decomposed in subspace base states, which finally processed by a CNN component for constructing the final representation.

\emph{Remark:} The above works outperformed various existing neural models at the time when the papers were published. However, more recent neural models, such as BERT based models~\cite{trec-qa-state-of-art, google-sst-state-of-art} have achieved a largely improved performance on the same tasks. It is worth exploring the integration of quantum models with the new BERT architecture~\cite{devlin2019bert} in the future.

Readers interested in tensor networks for representation learning, with or without quantum-inspirations, can also refer to \cite{zhang2019generalized_lm_tensor} which introduces a Tensor Space Language model (TSLM) by building higher dimensional tensors using word vectors, leading to a generalisation of n-gram models. 

In \cite{QINM_sigir20}, authors integrate quantum interference phenomena in neural networks with application to ad-hoc retrieval. Existing neural IR models are formulated in terms of classical probability. For example, if $q_i$ represent sub-units (e.q. words) of a query $Q$, then neural models first perform sub-unit level matching and then aggregate the scores obtained to calculate a final relevance score. Formulated in terms of probabilities, such aggregation follows the law of total probability:
\vspace{-2mm}
\begin{equation}
    P(R_D|Q) = P(q_1)P(R_D|q_1) + P(q_2)P(R_D|q_2)
\end{equation}

The authors hypothesise that a quantum-like cognitive interference can occur such that the aggregation of relevance scores can happen non-linearly, due to negative contribution of certain query or document sub-units. Thus the above equation becomes similar to Equation~\ref{interference-term}:
\vspace{-2mm}
\begin{equation}
        P(R_D|Q) = P(q_1)P(R_D|q_1) + P(q_2)P(R_D|q_2) + Int(R_D,Q,q_1,q_2)
\end{equation}
They proceed to model this interference term and incorporate into a neural architecture. Query and document states are represented as a superposition of their respective sub-unit states with the coefficients of document sub-unit states being the tf-idf values, and those of query sub-units being trainable parameters. A composite system is constructed by taking the tensor product of the query and document state vectors. Then, calculating relevance probability using the trace rule breaks down the probability into two parts -  similarity matching as used in neural IR models and the interference term, which is determined by the interaction between different document features. The similarity matching is achieved using a query attention mechanism, and the interference is modelled as a n-gram window convolution network. The new model is tested on Robust04 and Clueweb09-cat-B collections. While it performs better than various existing neural ranking models and even a vanilla-BERT~\cite{cedr-state-of-art-ir} (on P@20 metric on Robust04 dataset), it under-performs the current-state-of-the-art model~\cite{cedr-state-of-art-ir} on both NDCG@20 and P@20 on the Robust04. On the other hand, on the Clueweb-09-cat-B dataset it beats existing neural IR models and also the state-of-the-art XLNet model~\cite{yang2019xlnet} in NDCG@20 and ERR@20 metrics.
\\
\\
\textit{Quantum-inspired Classification}
\\
Classification algorithms inspired by Quantum detection theory~\cite{Helstrom1969} are discussed in \cite{DiBuccio2018, Tiwari:2018:TQF:3269206.3269304}. Binary classification is formulated as a signal detection problem.  Projectors are defined to detect a signal, i.e. whether a document belongs to a topic or not. The average cost of incorrect detection (false alarm - detecting a signal when it is absent, miss - failing to detect a signal) is expressed in terms of projection operators and is minimized over a training set. Experiments performed over Reuters Newswire dataset show comparable results with Naive Bayes and SVM algorithms. 
In \cite{DBLP:conf/lwa/JaiswalHFL18}, which used dimensionality reduction techniques for word embeddings, the computation times for the complex embeddings in \cite{DBLP:conf/rep4nlp/LiUWS18} is reduced, with additional application to the TREC-10 Question Classification task. 

\subsection{User Interactions}

\subsubsection{Projection Models}

The area of user interactions in IR has many sub-aspects, primarily - the cognitive level of interaction, understanding the user IN by reformulation, expansion of queries, and building a user profile based on historical interactions. Earlier, we mentioned about the work in \cite{Melucci2005CanVS,MelucciCMD}, which use multiple basis of a Hilbert space to model different user contexts. This work is further extended in \cite{Melucci2008,Melucci2007} to combine different user interaction and contextual features for Implicit Relevance Feedback (IRF). Their model uses interaction features like document display time, document saving, document bookmarking, webpage scrolling, webpage depth and document access frequency to construct a user interest profile. A basis vector represents each of these interaction features. The matching of documents against a user profile is done by projecting a document vector onto the subspace spanned by the basis vectors for the user profile. A large projection signifies high relevance of the document to the profile. The features described above are calculated for each document that the user has interacted with and a document-features correlation matrix is formed. Singular Value Decomposition (SVD) is performed to get the eigenvectors, which form the basis for a user profile. 

In \cite{Piwowarski2009,Frommholz2011}, a general framework for query reformulation using Quantum probabilities is described. The queries are represented as density matrices in a term space and query reformulation updates the query density matrix, which can be used to detect change in user IN in a search session. 

\begin{figure}
\centering
    \begin{subfigure}[t]{0.33\textwidth}
        \includegraphics[width=\textwidth]{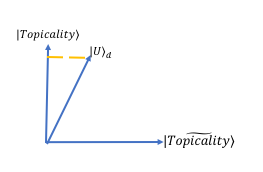}
        \caption{Topicality dimension in a single basis}
        \label{hilbert-space-a}

    \end{subfigure}
    \begin{subfigure}[t]{0.42\textwidth}
        \includegraphics[width=\textwidth]{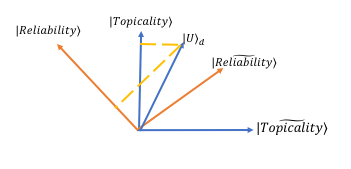}
         \caption{Topicality and Reliability in multiple basis}
         \label{hilbert-space-b}

  \end{subfigure}
  \vspace{-2mm}
\caption{Modelling user's perception of relevance dimensions in Hilbert space}
\label{hilbert-space-user-perception}
\vspace{-5mm}
\end{figure}

A Hilbert Space for user's perception of document relevance is constructed in~\cite{Uprety:2018:MMU:3209978.3210130}. It deals with the challenge of modeling multidimensional relevance of documents. In an extended Multidimensional User Relevance Model (MURM)~\cite{MURM-psychometrics, ASI:Jingfei}, seven factors or dimensions of relevance are identified, which influence user's judgment of a document. They are Topicality, Interest, Novelty, Understandability, Scope, Habit and Reliability. The features defined for each of these seven dimensions in \cite{ASI:Jingfei} are extracted from the retrieved documents of a query and fed into the LambdaMART~\cite{Burges2010FromRT} Learning to Rank algorithm. Thus for each relevance dimension, a document has a relevance score. In other words, for a query one gets seven different ranking lists, one for each dimension. The scores obtained are converted into probabilities using max-min normalization. Representing user's relevance perception of a document with respect to each relevance dimension as a vector, and perception of non-relevance as its orthogonal vector, a document can be represented in a two dimensional vector space~\cite{Uprety:2018:MMU:3209978.3210130}. The two relevance and non-relevance vectors for a relevance dimension form an orthonormal basis of the vector space. Figure \ref{hilbert-space-a} denotes one such vector for a user's perception of document relevance $\ket{U}_d$ with respect to Topicality dimension of relevance. The projection of the document perception vector on the Topicality vector is proportional to the probability of relevance of the document with respect to the Topicality dimension. Relevance and non-relevance of the document with respect to another dimension is represented as another set of orthogonal vectors, which, in general form another basis in the same Hilbert Space. In Figure \ref{hilbert-space-b}, we see that the projection of the document perception vector is different in the Reliability basis, suggesting that, while the document has a high relevance when considering the Topicality dimension, it is not so much relevant when considering the Reliability dimension.

The most popular interpretation of QT is that the state vector collapses upon measurement from a superposition to a definite state. When drawing an analogy, the user's cognitive state is generally considered as in a superposition of various Information Needs (IN) and on judging a document as relevant, it collapses to one particular IN. However, in practice, this may not always be the case. Even after judging documents, a user may still be in an ambiguous or superposed state and there may not have been an apparent change in the IN. The standard interpretation of state collapse may not accurately capture the evolution of IN. This challenge is investigated in \cite{quantum_weak_measurement} using a Quantum Weak Measurement (QWM) model. It is a generalization of the standard collapse model where the variance of measurement results is much larger. To test the weak measurement phenomena in user judgments, a study is carried out. Users are asked to judge documents on a -4 to 4 scale of relevance. For some query-document pairs, the users are asked to judge the same document for a second time. According to the standard collapse of the IN, after the judging a document as relevant, subsequent judgments will produce the same relevance result. However, it is found that in many cases, users change the relevance decisions. This will happen only when after the judging the document for the first time, the users were still uncertain about the document and their IN. The user's cognitive state was still superposed. This is especially the case where judgments on some documents are not trivial and difficult to make. The weak measurement model involves the Two State Vector Formalism (TSVF) of QT. In TSVF, the state of a system is not represented by a single vector $\ket{\psi}$ but rather by two vectors $\ket{\phi}$ and $\ket{\psi}$, where one vector represents the state of the system in the past (relative to a time $t$), and the other represents the state after time $t$. A weak measurement of an observable $W$ on the system is given by: 
\vspace{-2mm}
\begin{equation} \label{weak_measurement}
    w = \frac{\bra{\phi}W\ket{\psi}}{\braket{\phi}{\psi}}
\end{equation}

This type of Quantum measurement is applied in case of session search. A user's IN is represented by two vectors. One contains the historical session information in terms of Implicit relevance feedback, and the other represents the current query, in terms of Pseudo Relevance Feedback. The document vectors, $\ket{d_i}$, are calculated using word embedding techniques, and then the corresponding projectors $P_{D_i} = \ket{d_i}\bra{d_i}$ are constructed. The relevance probability of document $d_i$ using weak measurement is calculated as:
\vspace{-2mm}
\begin{equation}
    p = \frac{\bra{\phi_{past}}\ket{d_i}\bra{d_i}\ket{\phi_{curr}}}{\braket{\phi_{past}}{\phi_{curr}}}
\end{equation}

The experiment is conducted on the TREC Clueweb12 document collection and the QWM method gives a better performance than the Quantum Language Model and its variations, as well as some state-of-the-art classical IR models.

The TSVF is also used in \cite{two_state_vector} to modify the query density matrix in QLM. A quantum state is denoted as $\bra{\psi_{post}} \ket{\psi_{pre}}$. Here $\ket{\psi_{pre}}$ is a state evolving from the past to the present and $\bra{\psi_{post}}$ is a state devolving from future to the present. The previous user query in the session is considered for the past state and the current query for which the documents are to be retrieved is considered as the future query. Then, separate projectors are constructed for the past and future queries and the density matrix $\rho_d$ for the document is estimated in the following manner:
\vspace{-2mm}
\begin{equation}
    \rho_d = \arg\max(\sum_{i=1}^{M_{past}}\log tr(\rho_d\Pi_i) + \sum_{j=1}^{M_{future}}\log tr(\rho_d\Pi_j)) 
\end{equation}

where $M_{past}$ and $M_{future}$ are the number of projectors (made up of single terms or compound terms) in the past query and the future query respectively.

%In comparison with the QLM based query expansion model \cite{Li:2018:QLM:3234944.3234970} discussed in the previous subsection, this QWM-based QLM performs slightly less well despite it makes use of some user interaction data. One possible reason is that ...  

\subsubsection{Feedback}

The query drift problem presented in the previous subsection is approached using user's search history in \cite{Zhang2016AQQ}. A document is represented as a superposition of query vectors for current query and for a latent query defined by the user's query history.
\vspace{-5mm}
\begin{align}
\ket{d} = a_d\ket{q_c} + b_d\ket{q_h}
\end{align}
$q_h$ denotes the user IN that the user implicitly has in mind based on historical context, but has not been explicitly expressed into words. A document, in the superposition state of being relevant to both the current ($q_c$) and latent query ($q_h$), is then evaluated in terms of an expanded query. This is similar to the double slit experiment analogy with the two slits representing $q_c$ and $q_h$ and the detector screen representing the evaluation of this document in terms of the expanded query. Thus the document relevance with respect to the queries $q_c$ and $q_h$ interfere with each other. If $\ket{q_e}$ represents the vector for the expanded query and $\ket{d} = a_d\ket{q_c} + b_d\ket{q_h}$, then the projection of document onto the expanded query vector is:
\vspace{-3mm}
\begin{align}
d \rightarrow q_e &= |\bra{q_e}\ket{d}|^2  \\ \nonumber
&= |a_d\bra{q_e}\ket{q_c} + b_d\bra{q_e}\ket{q_h}|^2 \\ \nonumber
&= |\bra{q_c}\ket{d}\bra{q_e}\ket{q_c} + \bra{q_h}\ket{d}\bra{q_e}\ket{q_h}|^2 \\ \nonumber
&= |\bra{q_c}\ket{d}\bra{q_e}\ket{q_c}|^2 + |\bra{q_h}\ket{d}\bra{q_e}\ket{q_h}|^2 + 2\bra{q_c}\ket{d}\bra{q_e}\ket{q_c}\bra{q_h}\ket{d}\bra{q_e}\ket{q_h}\cos\theta
\end{align}
where $\theta$ is the phase between the two paths $d\rightarrow q_c \rightarrow q_e$ and $d\rightarrow q_h \rightarrow q_e$. We get interference between the two paths, because the actual path is superposed, $d\rightarrow (q_c \& q_h) \rightarrow q_e$. The first round retrieval is assumed to be using both the current and the latent query at the same time. This method of query expansion using user's previous interactions, is termed as the Quantum Query Expansion (QQE) approach for session search. It gives better results than the QFM discussed in the previous subsection, over the NDCG evaluation measure. 
\cite{QiuchiLi2015ModelingMR} propose a Session-QLM (SQLM) to model the evolving nature of user's IN in a search session. The evolution is modelled using density matrix transformation. The density matrix is constructed using user interaction features like clicked and skipped documents, dwell time, click sequence, etc. User's historical queries in context is also used in~\cite{Li2016} for a Contextual Quantum Language Model (CQLM). The density matrices are constructed to represent the language models for the current query and for the historical queries in a search session. They are then combined to give the CQLM.
\vspace{-2mm}
\begin{align}
\rho_{CQLM} = \xi \times \rho_c + (1-\xi) \times \rho_h
\end{align}
where $\xi \in [0,1]$ combines the two language models. 

The construction of $\rho_h$ is done by combining all the $\rho_{h_i}$ of the previous queries in the session. The historical queries in the session which are similar to the current query are given more weight. Hence:
\vspace{-1mm}
\begin{equation}
\rho_h = \sum_{i=1}^{n-1}{\gamma_i \times \rho_{h_i}}
\end{equation}
where $\gamma_i$ is the similarity between current query $q_c$ and previous query $q_i$. The similarity is calculated by "representing each query as a TF-IDF vector, derived from the concatenation of all of its result documents."~\cite{Li2016}.

The CQLM, however, was not designed to capture the evolution of user IN. To address this issue, the same paper further proposed an Adaptive CQLM (ACQLM) to model the evolution of user IN. The basic idea is to decompose the current query into three - a common part, an added part and a removed part, relative to the previous queries in the session. For example, if $q_k = wxy$, $q_{k-1} = xyz$, ``then $xy$ is the common part, $z$ is the added part and $w$ is the removed part. The common part indicates the user's underlying search topic/theme for the session. The removed and added parts reflect the change in IN''~\cite{Li2016}. The ACQLM adjusts the QLM in such a way, as to assign a higher probability to the terms (or composite terms) of the common and added parts. Thus the ACQLM builds upon the CQLM by incorporating query change signals in a structured and intuitive way, moving the QLM into the right direction.

\subsubsection{Context Effects}
A series of research has been carried out from the user cognitive aspect of IR, drawing parallels from QT and using the Quantum framework to model and explain some of the aspects. An early work~\cite{junwang} investigated the interference in relevance judgment of a topic caused by another topic. Consider the topics ``Brave Heart'' (William Wallace's nickname and the name for his film biography) and ``William Wallace'', and a biographical article about William Wallace. Both topics are relevant to the article. Consider another topic about ``William Wallace's wife''. In a user study, it was found out that when the topics ``Brave Heart'' and ``William Wallace'' were displayed together for the article, $93\%$ users chose to judge the article as relevant to ``William Wallace'' and only $14\%$ chose it as being relevant to the topic ``Brave Heart''. However, when ``Brave Heart'' was displayed together with ``William Wallace's wife'', $89\%$ of the users judged ``Brave Heart'' as relevant to the article and $5\%$ judged ``William Wallace's wife'' to be relevant. There were experiments conducted with different topics and articles and such type of context effects were found, where the presence of one topic or document influences the relevance judgment of another topic or document. In the first case, ``William Wallace'' is highly relevant to the article. It sets a high comparison baseline which affects the judgment for the topic ``Brave Heart'' and results in a low probability of relevance. However, it appears more relevant in comparison with ``William Wallace's wife''. For a Quantum probabilistic explanation of this result, we regard ``William Wallace'' and ``William Wallace's Wife'' as two different contexts for the topic ``Brave Heart''. Each context is described by a basis. So a document or topic $d$ can be represented in the context basis as:
\vspace{-2mm}
\begin{equation}
\ket{d} = a_1\ket{q_1} + a_2\ket{\bar{q_1}}
\end{equation}
where $\ket{\bar{q_1}}$ represents the absence of context $q_1$. Representing a query $q$ in the same basis as $\ket{q} = b_1\ket{q_1} + b_2\ket{\bar{q_1}}$, we can calculate the relevance of the document $d$ for query $q$ as:
\vspace{-2mm}
\begin{align}
p(d|q) &= |\bra{q}\ket{d}|^2 \\ \nonumber
&= (a_1b_1+a_2b_2)*(a_1b_1+a_2b_2)^{\dagger} \\ \nonumber
&= a_1^2b_1^2 + a_2^2b_2^2 + 2a_1b_1a_2b_2\cos\theta
\end{align}
where the probability amplitudes are complex quantities, and $\theta$ represents the phase difference term. The third term is the interference term, which can be positive or negative depending upon the phase differences. For some contexts, the interference term is negative and the relevance of the same document for the query can be low, which explains why ``Brave Heart'' is judged less relevant when seen in the context of ``William Wallace''. There is a negative interference term that lowers the probability of relevance for the given query/article.

\begin{figure}
\centering
    \begin{subfigure}[t]{0.4\textwidth}
        \includegraphics[width=\textwidth]{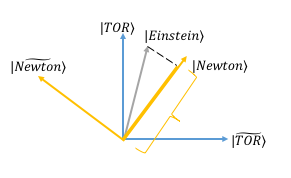}

    \end{subfigure}
    \begin{subfigure}[t]{0.4\textwidth}
        \includegraphics[width=\textwidth]{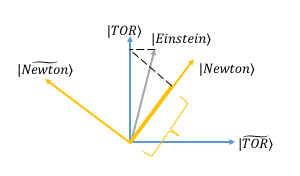}

  \end{subfigure}
  \vspace{-7mm}
\caption{On viewing document about Theory of Relativity, the judgment of topic Newton is lower for the query Einstein}
\label{order-effects-judgment}
\vspace{-5mm}
\end{figure} 

A follow-on work~\cite{ASI:Order} further explored the influence of context in document relevance judgment. It specifically investigates the presence of Order Effects in relevance judgment of documents. In the experiment, users are shown a pair of documents for a query and the relevance judgment by the user for a document is affected by the order, in which the document is presented. For example, for the query ``Albert Einstein'' users are shown documents about ``Issac Newton'' and ``Theory of Relativity''. The relevance probability of ``Issac Newton'' is lower when it is shown after ``Theory of Relativity'' (called a comparative context) than when it is shown first (non-comparative context). In simple terms, having seen a more relevant document first, user's perception of relevance for a particular document may be lower. This can be explained as an Order Effect due to incompatibility between the topics, as shown in Figure \ref{order-effects-judgment}. The paper also tested the Quantum Question Order inequality~\cite{Wang2013}, which is an inequality for testing incompatibility in decision making systems. 

One of the earliest works to investigate order effects in the different relevance dimensions is \cite{Bruza10.3389/fpsyg.2014.00612}. A user study was conducted, in which participants were asked questions about different pairs of relevance dimensions for a document, e.g. Credibility and Understandability, etc. It was found that the judgement of credibility, novelty, etc. was different depending upon the order in which they were asked to judge. 

\begin{figure}
\centering
    \begin{subfigure}[t]{0.35\textwidth}
        \includegraphics[width=\textwidth]{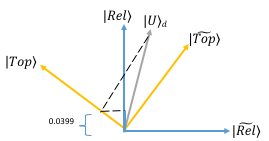}
        \caption{User considers the sequence Topicality,   \\ Reliability while judging first document}
        \label{order-effects-logs-a}
    \end{subfigure}
    \begin{subfigure}[t]{0.35\textwidth}
        \includegraphics[width=\textwidth]{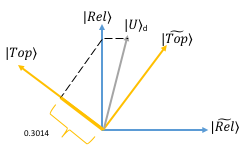}
         \caption{User considers the sequence Reliability, Topicality while judging second document}
        \label{order-effects-logs-b}

  \end{subfigure}
  \vspace{-3mm}
  \caption{Different order of consideration of dimensions leads to different final probability}
  \label{order-effects-logs}
  \vspace{-6mm}

\end{figure}

Similar order effects using query log data have been investigated in \cite{Uprety:2018:IOE:3234944.3234972}. The method of constructing a Hilbert space for multidimensional document relevance perception from \cite{Uprety:2018:MMU:3209978.3210130} is used (discussed earlier in this subsection). It is assumed that a user may consider multiple relevance dimensions while judging a document, for example, topicality and novelty. The relevance perception vectors corresponding to different relevance dimensions are in general incompatible in the Hilbert space representation. Thus different orders of consideration of the relevance dimensions may lead to different final judgment of the document. To investigate such behaviour in query log data, a subset of queries are found, where the top two retrieved documents have similar scores of relevance in all the seven dimensions. Yet the first document in the ranked list is not judged relevant, but the second one is. A small set of such queries are indeed found and order effects arising out of different order of consideration of relevance dimensions is offered as a possible explanation for such queries. Figure \ref{order-effects-logs} explains the order effect for two documents $d_1$ and $d_2$ (ranking order for a query), which have the exact same Hilbert space, yet only $d_2$ is clicked. For $d_1$, if the user first considers Topicality and then Reliability to judge document $d_1$, then the final probability of judgment obtained is $0.0399$ (Figure \ref{order-effects-logs-a}). Whereas, for $d_2$, if the order is reversed, the probability of final judgment obtained will be $0.3014$ (Figure~\ref{order-effects-logs-b}), much larger in this case. However, an important question is why the order is reversed in the user's mind for the next document. The authors argue that it could be due to a memory bias - the user can use the last relevance dimension considered for the previous document as the first dimension while judging the current document. Also, there is a possibility that such behaviour can be due to a variety of different reasons, or just random errors in the log data. Nonetheless, a quantum cognitive explanation based on order effects is a possibility. 

As we see that there is some evidence of incompatibility between different relevance dimensions, \cite{QI_bell} investigated the violation of Bell-type inequalities for multidimensional relevance judgment data. A violation of Bell-type inequalities would confirm the quantum nature of data. However, no such violation is found due to lacking probabilities of relevance available for joint judgment of a pair of documents in their dataset.

Order effect in risk and ambiguity is also investigated and observed in Information Foraging Theory in \cite{Wittek2016}. This paper identifies a theoretical limit to simultaneous consideration of risk and ambiguity in decision making using eye tracking data, analogous to the uncertainty principle.

In \cite{uprety2019modelling}, the phenomena of incompatibility and order effects between relevance dimensions has been studied through a novel protocol design inspired from the Stern-Gerlach experiment of Physics. For a query-document pair, two groups of users were asked three questions relating to Topicality (T), Understandability (U) and Reliability (R) of a document, in orders TUR and TRU respectively. A complex-valued Hilbert space for the user cognitive state is constructed using the data obtained from the experiment, which is used to construct operators corresponding to the T, U and R measurements/judgements. Interference and incompatibility is discussed using these operators. This is the first work where complex numbers are used to capture interactions between relevance dimensions such as incompatibility and interference. It is extended in \cite{Uprety2020ecir} to test the violation of a Kolmogorovian probability axiom:
\vspace{-2mm}
\begin{equation}\label{eq-kolmogorov}
     0=\delta=P(A \lor B)-P(A)-P(B)+P(A \land B)
        \vspace{0mm}
\end{equation}
where the events $A$ and $B$ are the questions regarding Understandability and Reliability of a document. The conjunction and disjunction questions are asked through a specific experiment design and a violation of the above equality is observed in the data. Quantum model predicts a violation for all queries. This paper also
compares quantum and Bayesian models for predicting  multidimensional relevance probabilities. Quantum predictions are consistently closer to the experimental data, while predictions from the Bayesian model deviate significantly in some cases.

\vspace{-2mm}
\section{Summary and Limitations}
\vspace{-1mm}
\begin{table}
\begin{tabular}{|p{2.25cm}|p{2.3cm}|p{11.5cm}|}
\hline
Area                                                                                   & Sub-area                                                                           & References \\ \hline
\multirow{5}{*}{\begin{tabular}[c]{@{}l@{}}Representation \\ and Ranking\end{tabular}} & Projection Models                                                                  &  \cite{Rijsbergen:2004:GIR:993731},\cite{ Melucci2005CanVS},\cite{MelucciCMD}, \cite{Piwowarski2009},\cite{Piwowarski:2008:SCR:1416950.1416951},\cite{Piwowarski:2010:QTB:1871437.1871450},\cite{ Piwowarski:2010:EMR:1937055.1937067},\cite{ Piwowarski2010FilteringDW}, \cite{queryalgebra}, \cite{Frommholz:2010:SPQ:1840784.1840802}          \\ \cline{2-3} 
                                                                                       & \begin{tabular}[c]{@{}l@{}}Quantum \\ Language Models\end{tabular}                 &  \cite{Sordoni:2013:MTD:2484028.2484098}, \cite{Xie2015ModelingQE},\cite{Zhang_many_body} \cite{Hou2009CharacterizingPH}, \cite{Hou:2013:MPH:2493175.2493177}, \cite{unsupervised_sentiment_analysis}, \cite{Zhang2018EndtoEndQL}, \cite{Zhang_many_body} , \cite{blacoe_quantum_distributional},\cite{BlacoeQI}, \cite{towards_QLM}, \cite{Li:2018:QLM:3234944.3234970}         \\ \cline{2-3} 
                                                                                       & \begin{tabular}[c]{@{}l@{}}Quantum-inspired\\  Ranking\end{tabular}                &  \cite{Zuccon2009}, \cite{Zuccon:2010:UQP:2128344.2128384}, \cite{Zhao2011}, \cite{Zhang2011InvestigatingQP}          \\ \cline{2-3} 
                                                                                       & Multimodal IR                                                                      & \cite{wang2010tensor}, \cite{kaliciak2011contextual},\cite{ kaliciak2013combining}, \cite{dimitris_QI}, \cite{zhang2018quantum}, \cite{zhang2020quantum}           \\ \cline{2-3} 
                                                                                       & \begin{tabular}[c]{@{}l@{}}Quantum-inspired\\ Representation \\ Learning\end{tabular} &
                                                                                       \cite{aerts_czachor}, \cite{BruzaC05}, \cite{Sordoni:2013:MLT:2505515.2507854}, \cite{10.1007/Guido_complex_num_ICTIR}, \cite{10.1007/complexIR_Wittek_Guido}, \cite{DBLP:conf/rep4nlp/LiUWS18}, \cite{DBLP:conf/lwa/JaiswalHFL18}, \cite{wang2019semantic}, \cite{li2019cnm}, \cite{DiBuccio2018}, \cite{Tiwari:2018:TQF:3269206.3269304}\\ \hline
\multirow{3}{*}{User Interactions}                                                     & Projection Models                                                                  &  \cite{Melucci2007}, \cite{Melucci2008}, \cite{Piwowarski2009}, \cite{Frommholz2011}, \cite{Uprety:2018:MMU:3209978.3210130}, \cite{quantum_weak_measurement}, \cite{two_state_vector}          \\ \cline{2-3} 
                                                                                       & Feedback                                                                           &   
                                                             \cite{QiuchiLi2015ModelingMR}, \cite{Zhang2016AQQ}, \cite{Li2016}
                             \\\cline{2-3} 
                                                                                       & Context Effects                                                                    &   \cite{junwang}, \cite{ASI:Order}, \cite{Uprety:2018:IOE:3234944.3234972}, \cite{QI_bell}, \cite{Wittek2016}, \cite{uprety2019modelling}, \cite{Uprety2020ecir}         \\ \hline
\begin{tabular}[c]{@{}l@{}}Quantum-inspired\\Neural Networks \end{tabular}  & - & \cite{Zhang2018EndtoEndQL}, \cite{DBLP:conf/rep4nlp/LiUWS18}, \cite{DBLP:conf/lwa/JaiswalHFL18}, \cite{Zhang_many_body},
\cite{zhang2019quantum} \cite{zhang2020quantum}, \cite{wang2019semantic}, \cite{li2019cnm}, \cite{zhang2019generalized_lm_tensor}, \cite{QINM_sigir20}\\
\hline
\end{tabular} \caption{Review Summary}
\label{table-summary}
\vspace{-11mm}
\end{table}

van Rijsbergen's seminal work introduces us to similarities between the mathematical representation of microscopic particles in QT and information objects in IR. It does not delve into much depth over the distinct advantages of the quantum framework over traditional IR frameworks.

Early research inspired by van Rijsbergen's ideas implement QT-based ad-hoc IR models by considering information need space as Hilbert space and introducing ideas of superposition for ambiguous queries. These representations provide a good starting point in QR, but they generally fail to outperform the state-of-art methods in IR.

The Quantum Language Model (QLM) is a promising application and intends to solve a crucial problem in NLP and IR - of representing compound terms in relation to the individual terms. Superposition principle is made use of and a quantum algorithm to build a language model is applied. It performs better than baseline models like tf-idf and BM25. The later quantum-inspired language models show marked improvement over the QLM but need to be applied on a wide variety of IR and NLP tasks and compared with the state-of-the-art baselines. The complex word embedding is another promising approach, however there is a lack of clarity as to why this methods performs better than some classical methods and what is the intuition behind the interference terms and complex phases. 

The Quantum Probability Ranking Principle is an important milestone in quantum-inspired IR as it approaches and combines QT and IR from an axiomatic point of view. However, the problem of quantifying the interference term remains and document similarity approaches applied do not take the quantum advantage. One needs to devise a way to subscribe complex phases to documents and then calculate the interference terms. 

The query fusion and query expansion approaches make use of superposition and interference phenomena, however it is difficult to get an intuitive explanation of how these two are coming into effect and providing the advantage over classical methods. The Contextual QLM (CQLM) and Adaptive-CQLM are promising applications of the QLM to incorporate user interactions, however they are outperformed by the state-of-art machine learning based methods.

The integration of the quantum framework to neural networks is promising and combines the representational complexity of neural networks with the probabilistic generalization provided by the quantum framework, especially when complex numbers are included. However, as we see in the results reported in this survey, the state-of-the-art neural networks outperform quantum-inspired models. A reason could be that the datasets used are mostly static, devoid of context, the human factor and its complexities, but it is not the case in real applications. 

The cognitive experiments on order effects in document judgment provide a good insight into why quantum probability is useful in modeling human decision making. However, most of these experiments are only performed on small user collected samples and need to be conducted on real world search data. Also, they do not yet provide a way to make use of the order effect information to improve the effectiveness of IR systems. 

We summarise the survey in Table \ref{table-summary} with the papers categorised into the sub-areas of IR mentioned in Figure \ref{quantum_ir_brief}. We also list papers which use quantum-inspired neural networks, encompassing all the other sub-areas.

\vspace{-2.5mm}
\section{Future Directions in Quantum-inspired IR}
\vspace{-1mm}
Quantum Theory was developed as a framework to explain the counter-intuitive behaviour of microscopic particles which could not be explained using traditional probability and logic models. Hence, the most important thing to consider while applying the quantum framework to IR or any other computational sciences, is the existence of such non-classical data (which violates classical logic, e.g. Boolean logic). There is substantial evidence in behavioural sciences that data obtained from human information interaction and decision-making can be quantum-like data. Phenomena like conjunction and disjunction fallacy, violation of law of total probability (LTP), similarity effects \cite{Tversky1977-similarity}, etc. can be investigated in user behaviour in IR. Conjunction and disjunction fallacies can exist in relevance judgment of documents. Although the QPRP incorporates the interference term in document ranking, it does not explicitly occur due to the violation of LTP. In fact, LTP violation can be investigated in IR when users make decisions under ambiguity and the cognitive state can be modelled as a superposition of different states.

Hence, the presence of quantum-like data and quantum phenomena in IR can completely change our understanding of the fundamentals of IR. A strong candidate to start with is to rethink the concept of relevance and the interpretation of probability of relevance. In the Cranfield methodology of building IR test collections, a common practice is to reject annotator disagreement as noise and fix one relevance label for a given document which has been judged by the majority of annotators. However, it is possible that both labels exist for the document for the same query, and the disagreement between annotators is due to their different contexts which leads to different semantic interpretations of the document. Or, there can be ambiguity in the document content and meta-data (e.g. arising out of its lack of credibility, novelty, etc.) which can put annotators in two minds, leading to different relevance labels. Hence a document needs to be modelled as both relevant and non-relevant for a query. We see that QT has tools to model two mutually exclusive states of a system as a single state (Superposition). Future work can begin with constructing such contextual datasets~\cite{inel2014crowdtruth}. 

With such contextual data in hand, subspace generalization of vector spaces can be utilized to create more contextual and dynamic representations. Representing a document or a word with a subspace rather than a vector allows for representing its different aspects in the vectors which span the subspace. This can be instrumental in capturing meaning in the data in the same way as humans do (subject to multiple contexts, arising out of external factors or intrinsic cognitive biases). Integrating different Hilbert Spaces by using the tensor product is a useful technique to fuse different modes of information like text, images, audio, etc. as in multi-modal IR. It can also be useful in fusing different cognitive aspects of information (Polyrepresention) e.g., product reviews sentiment and brand credibility.

The presence of incompatibility in judgments can render many user models, which assume joint distributions between variables, as ineffective. For example, in~\cite{joint_sentiment_topic_model}, a joint sentiment-topic model to detect sentiment and topics simultaneously from text is proposed. The joint probability of a word, topic and sentiment label assigned is written as $ p(w, z, l) = p(w|z,l)p(z,l)$
where $w$ is a word, $z$ is a topic of the document and $l$ is the sentiment label of the document. If there is incompatibility between the sentiment and topic of a document, then the predictions based on this model would not match the user's decisions \cite{Bruza10.3389/fpsyg.2014.00612}. This is something which can be captured using the quantum framework and quantum-inspired models will be better equipped to handle such cases. 

This research can further benefit from a formal model of quantum-inspired neural networks with a theoretical basis and where the interference term and complex numbers naturally occur in the neural computations. We believe that best way forward in this field will be the integration of concepts and constructs from QT with the state-of-the-art machine learning models, e.g. neural networks. QT framework is best positioned to model human decisions under ambiguity and dynamic changes of context. Their fusion with QT can enhance their ability to model complex human behavioural data, building a platform for more human-centred Artificial Intelligence. It is known that the neural models suffer from the problem of generalisation. One way to tackle the problem is that multiple statistical hypotheses of a neural network can be preserved in the form of a superposition state in the middle layer and used by the model decision layer. Traditionally, a deep learning model is based on one hypothesis, thus limiting the model’s generalisation ability. The output layer of the deep learning model can integrate multiple statistical hypotheses that the hidden layer retained and learned during training. The nature of reserving multi-hypothesis at the same time is like quantum superposition. This makes for some exciting research prospects in the future.

The authors are hopeful that this literature survey is able to provide a clear picture of the quantum-inspired IR field and set a road-map for researchers to take this field forward.
\vspace{-3mm}
\bibliography{bibliography}
\bibliographystyle{ACM-Reference-Format}

\end{document}